\documentclass[11pt, a4paper]{article}

\usepackage{etoolbox}
\usepackage[yyyymmdd]{datetime}

\usepackage[mathscr]{eucal}
\usepackage{amsmath}
\usepackage{amssymb}
\usepackage{mathtools}
\usepackage{wasysym}
\usepackage{bbm}

\usepackage[latin1]{inputenc}
\usepackage[T1]{fontenc}
\usepackage{url}
\usepackage{fancyhdr}
\usepackage{moresize}

\usepackage{enumerate}
\usepackage[shortlabels]{enumitem}

\usepackage{geometry}
\usepackage{lscape}
\usepackage{soul}
\usepackage{lastpage}
\usepackage{setspace}
\usepackage{hyperref}
\usepackage[bottom,hang,flushmargin]{footmisc}
\usepackage{adjustbox}

\usepackage{titlesec}
\usepackage{placeins}

\usepackage{graphicx}
\usepackage{subcaption}
\usepackage{float}
\usepackage{rotating}

\usepackage{tabularx}
\usepackage{dcolumn}
\usepackage{multirow}

\usepackage{tikz}
\usepackage{xcolor}
\usepackage{empheq}
\usepackage[most]{tcolorbox}

\usepackage[round]{natbib} 

\fancypagestyle{plain}{  \fancyhf{}
	\rfoot{\thepage}
	
	}

\renewcommand\thesection{\arabic{section}}
\titleformat{\section}{\bfseries\fontsize{12pt}{12}}{\thesection.}{0.8em}{}

\renewcommand\thesubsection{\arabic{section}.\arabic{subsection}}
\titleformat{\subsection}{\bfseries\fontsize{12pt}{12}}{\thesubsection \hspace{8pt}}{0em}{}
\titlespacing\subsection{0pt}{8pt plus 2pt minus 2pt}{4pt plus 2pt minus 2pt}

\renewcommand\thesubsubsection{\arabic{section}.\arabic{subsection}.\arabic{subsubsection}}
\titleformat{\subsubsection}{\bfseries\fontsize{11pt}{11}}{\thesubsubsection \hspace{8pt}}{0em}{}

\setlist[itemize]{itemsep=-3pt, topsep=-5pt, label=-, after=\vspace{7pt}}
\setlist[enumerate]{itemsep=-3pt, topsep=-5pt, after=\vspace{7pt}}

\makeatletter
\def\hlinewd#1{%
	\noalign{\ifnum0=`}\fi\hrule \@height #1 %
	\futurelet\reserved@a\@xhline}
\makeatother

\counterwithin{table}{section}
\renewcommand\thetable{\arabic{section}.\arabic{table}}   

\newcolumntype{L}[1]{>{\raggedright\let\newline\\\arraybackslash\hspace{0pt}}m{#1}}
\newcolumntype{C}[1]{>{\centering\let\newline\\\arraybackslash\hspace{0pt}}m{#1}}
\newcolumntype{R}[1]{>{\raggedleft\let\newline\\\arraybackslash\hspace{0pt}}m{#1}}

\renewcommand\arraystretch{1.15}

\usepackage{chngcntr}
\counterwithin{figure}{section}
\renewcommand{\thefigure}{\arabic{section}.\arabic{figure}}

\newcommand\ftnotetable[1]{\captionsetup{size=scriptsize, font = normal, justification=justified,skip=0pt}\vspace{-8pt}\caption*{#1} }
\newcommand\ftnotefigure[1]{\captionsetup{size=scriptsize, font = normal, justification=justified}\caption*{#1}}

\numberwithin{equation}{section}

\definecolor{bluecite}{RGB}{60, 100, 180}
\hypersetup{
	colorlinks = true,
	citecolor = bluecite,
	linkcolor = bluecite
}

\makeatletter
\AtBeginDocument{%
	\let\plain@equationautorefname\equationautorefname
	\def\equationautorefname{\plain@equationautorefname\@autoref@insert@tagform}%
	\def\@autoref@insert@tagform~#1\null{~(#1\null)}%
}
\makeatother

\usepackage{titling}

\newcommand\blfootnote[1]{%
	\begingroup
	\renewcommand\thefootnote{}\footnote{#1}%
	\addtocounter{footnote}{-1}%
	\endgroup
}
	
 \usetikzlibrary{shapes,arrows}
 \usetikzlibrary{positioning,calc}
 
 \newtcolorbox{Box1}[2][]{
 	lower separated=false,
 	colback=white!95!gray,
 	colframe=white, fonttitle=\bfseries,
 	colbacktitle=white!80!gray,
 	coltitle=black,
 	enhanced,
 	attach boxed title to top left={xshift=0.5cm,yshift=-2mm},
 	title=#2,#1}
 
\usepackage{datatool}
\DTLsetseparator{|}
\DTLloaddb{DynamicWriting1}{./Output/DynamicWriting/1_2_simulation_1.csv}
\DTLloaddb{DynamicWriting2}{./Output/DynamicWriting/1_1_simulation_2.csv}

 \title{
 	Estimating Duration Dependence in Job Search: \\ the Within-Estimation Duration Bias
}
\author{Jeremy Zuchuat \thanks{Contact: jeremy.zuchuat@unil.ch} \\ \textit{University of Lausanne}}

\begin{document}
\begin{filecontents*}{./Output/DynamicWriting/1_2_simulation_1.csv}
object|value
K|10,000
N|1,000
\end{filecontents*}

\begin{filecontents*}{./Output/DynamicWriting/1_1_simulation_2.csv}
object|value
K|5,000
N|500
\end{filecontents*}



\maketitle

\thispagestyle{empty}

\begin{center}
	\vspace{-30pt}
	\centering\textsc{Working paper}
\end{center}

\begin{abstract}
	\footnotesize
	\noindent
	Many recent studies use individual longitudinal data to analyze job search behaviors. 
 	Such data allow the use of fixed-effects models, which supposedly address the issue of dynamic selection and make it possible to identify the structural effect of time. 
 	However, using fixed effects can induce a sizable within-estimation bias if job search outcomes take specific values at the time job seekers exit unemployment. 
 	 This pattern creates an undesirable mechanical correlation between the error term and the time regressor. 
 	  This paper derives the conditions under which the fixed-effects estimator provides valid estimates of structural duration-dependence relationships. 
 	   Using Monte Carlo simulations, we show that the magnitude of the bias can be extremely large. 
 	   Our results are not limited to the job search context but naturally extend to any framework in which longitudinal data are used to measure the structural effect of time.

\end{abstract}

\blfootnote{
	I am grateful to Rafael Lalive, Aderonke Osikominu, Nicola Mauri and Josef Zweimüller for their helpful guidance all along the writing of this paper. 
}

\textbf{Keywords}: Job search dynamics, duration dependence, fixed effects, within-estimation, bias, attrition.\\[5pt]
\textbf{JEL}: C23, C41, J64

\clearpage
	
\setcounter{page}{1}



\section{Introduction}

%
%

Negative duration dependence in job finding has long been documented: the longer job seekers remain unemployed, the lower their instantaneous probability of finding a job \citep{kaitz1970analyzing}.  
Conceptually, this phenomenon arises from two coexisting forces.  
On the one hand, empirically observed data are subject to dynamic selection: due to sample attrition, the average job seeker observed in later stages of unemployment has lower employment prospects than the average unemployed observed in the early phases.  
On the other hand, elapsed unemployment duration structurally affects the job search process through its influence on workers' and firms' behaviors (e.g., discouragement, discrimination against the long-term unemployed).

%
%
Over the past decades, numerous studies have sought to disentangle the structural component of duration dependence from its dynamically selective aspect.  
A variety of methodologies have been employed, ranging from purely statistical approaches (\citealp{heckman1984method,van2001duration,alvarez2016decomposing}) to experimental designs (\citealp{oberholzer2008nonemployment, kroft2013duration, eriksson2014employers, farber2016determinants}).  
More recently, researchers have taken advantage of richer datasets to uncover structural duration dependence patterns in labor markets.  
These datasets typically consist of longitudinal records of individuals' job search activity, collected through online search platforms (\citealp{kudlyak2013systematic,faberman2019intensity, marinescu2021unemployment}), periodic surveys (\citealp{belot2019providing,dellavigna2022evidence}), or job search diaries maintained at public employment services (\citealp{fluchtmann2021dynamics,lalive2022ddjs, cederlof2025duration}).

%
%

These data are assumed to allow researchers to account carefully for dynamic selection through the use of fixed effects models.  
This class of models is believed to provide unbiased estimates of structural duration dependence profiles, net of dynamic selection related to workers' productivity or employment history.  
They have notably been applied to measure duration dependence in job search effort (\citealp{faberman2019intensity,fluchtmann2021dynamics,marinescu2021unemployment,lalive2022ddjs}), in the quality of job applications (\citealp{kudlyak2013systematic}) or in job interviews (\citealp{cederlof2025duration}) .
Renowned for their robustness, fixed effects models have been applied to estimate structural duration dependence relationships, often without considering the specific nature of the independent variable of interest, time itself.

%
%

This paper demonstrates that, in certain cases, the within-transformation applied by the fixed effects estimator can induce a spurious mechanical correlation between the error term and the time regressor.  
This can lead to substantial bias in the structural duration profile estimated using the fixed effects approach.  
Using Monte Carlo simulations, we show that the magnitude of this \textit{within-estimation duration bias} can be extremely large, particularly when the dependent variable is related to the sample attrition mechanism.  
These findings are not limited to the context of job search, but extend naturally to any setting in which researchers aim to measure structural duration dependence using longitudinal data.

%
%

The remainder of the paper is organized as follows.  
\autoref{section:bias} derives the \textit{within-estimation duration bias} and identifies the conditions under which it may arise.  
\autoref{section:sim2} presents Monte Carlo simulations inspired by a real-world job search process, designed to quantify the presence and magnitude of the bias.  
\autoref{section:conclusion} concludes.



\section{The within-estimation duration bias}

\label{section:bias}

We first derive the conditions under which the fixed effects estimator applied to longitudinal data provides unbiased estimates of structural duration profiles.  
We then discuss the situations in which the \textit{within-estimation duration bias} is likely to arise.

\subsection{Within-estimaton of duration dependence}

We are interested in measuring the structural effect of time on a time-varying variable $y_{it}$, where $i$ denotes individuals (or spells) and $t$ denotes duration.  
In a job search context, this variable may represent, for example, the search effort of unemployed individuals (\citealp{faberman2019intensity, fluchtmann2021dynamics, marinescu2021unemployment, dellavigna2022evidence}) or the responses of firms (\citealp{kroft2013duration, eriksson2014employers, farber2016determinants, lalive2022ddjs, cederlof2025duration}).  
The data-generating process of $y_{it}$ is defined as
\begin{align}
	y_{it} =  \alpha_i + \beta t + \varepsilon_{it}
\end{align}
The value of $y_{it}$ depends on individual heterogeneity (through $\alpha_i$) and is structurally affected by duration (through $\beta t$). 
$\varepsilon_{it}$ represents an idiosyncratic error term. 

We assume access to individual longitudinal data.  
Each sampled individual is observed at different durations $t \in \{1,2,\ldots, T_i\}$.  
The total number of observation periods is individual-specific and given by $T_i = \min\{\tau_i, \bar{\tau}\}$.  
Here, $\tau_i$ represents the length of the spell, \textit{i.e.}, the number of periods after which individual $i$ exits the population of interest, while $\bar{\tau}$ denotes the maximum number of observation periods for all sampled individuals, \textit{i.e.}, the right-censoring duration.

Variability in $T_i$ is driven by sample attrition: some individuals exit the sample during the observation period ($T_i = \tau_i \leq \bar{\tau}$), while others have right-censored spells ($T_i = \bar{\tau} < \tau_i$).

As a result of individual heterogeneity, the observed sample may be subject to dynamic selection: depending on their $\alpha_i$, some individuals are more likely to exit the sample in early periods, and vice versa.\footnote{
	This does not apply if the attrition mechanism is unrelated to individual heterogeneity.
}  

A common approach to estimating parameters in the presence of individual heterogeneity is to rely on fixed effects.  
We adopt this approach and estimate the duration dependence parameter $\beta$ using a linear fixed effects model.  
Applying the standard within-transformation, we rewrite the previous equation as a within-individual specification:

\begin{align}
	\label{eq:within_specification}
	\tilde y_{it} &=  \beta \tilde t_i + \tilde \varepsilon_{it},
\end{align}
where $\tilde y_{it}  = y_{it}-\bar y_i$, 
$ \tilde t_i  =  t - \bar t_i$ and 
$\tilde \varepsilon_{it} = \varepsilon_{it} - \bar\varepsilon_i$.
Individual-specific averages are defined as 
$\bar y_i = \frac{1}{T_i}\sum^{T_i}_{t=1} y_{it}$, 
$\bar t_i= \frac{1}{T_i}\sum^{T_i}_{t=1} t$ and 
$\bar \varepsilon_i = \frac{1}{T_i}\sum^{T_i}_{t=1} \varepsilon_{it}$

The OLS estimator of $\beta$ from the above equation corresponds to the fixed effects estimator of the structural duration profile of $y_{it}$, denoted $\hat{\beta}_{FE}$.  
By substituting the data-generating process into the estimator equation and noting that $\bar{\tilde{t}} = 0$, we obtain:

\begin{equation}
	\label{equation:FE_estimator}
	\hat \beta_{FE} 
	= \frac{\sum^N_i \sum^{T_i}_t (\tilde y_{it} -  \bar{\tilde y}) (\tilde t_{i} -  \bar{\tilde t})}{\sum^N_i \sum^{T_i}_t (\tilde t_{i} -  \bar{\tilde t})^2}
	= \beta + 	\frac{\sum^N_i \sum^{T_i}_t  (\tilde \varepsilon_{it} -  \bar{\tilde \varepsilon}) \tilde t_{i}}{\sum^N_i \sum^{T_i}_t \tilde t_i^2}
\end{equation}

The expected value of the fixed effects estimator is equal to
\begin{equation}
	\mathbb{E}\left(\hat \beta_{FE} \right) 
	= \beta + \mathbb{E}\left(\frac{\sum^N_i \sum^{T_i}_t  (\tilde \varepsilon_{it} -  \bar{\tilde \varepsilon}) \tilde t_{i}}{\sum^N_i \sum^{T_i}_t \tilde t_{i}^2} \right)
\end{equation}

Assuming $\mathbb{E}(\tilde{\varepsilon}_{it} \mid \tilde{t}_i) = 0$, we obtain the standard result for the unbiasedness of the OLS (and fixed effects) estimator:\footnote{
	More detailed derivations of the estimator and the bias are provided in \autoref{section:theory_appendix}.
}\textsuperscript{,}\footnote{
	As a corollary of this assumption, unbiasedness requires that there is no within-individual correlation between the error term $\tilde{\varepsilon}_{it}$ and time $\tilde{t}_i$.
}

\begin{align}
	\mathbb{E}\left(\hat \beta_{FE} \right) 
	&= 
	\beta +  \mathbb{E}_i\left( \mathbb{E}\left(\frac{\sum^N_i \sum^{T_i}_t  (\tilde \varepsilon_{it}-\bar{\tilde \varepsilon}) \tilde t_{i}}{\sum^N_i \sum^{T_i}_t \tilde t_{i}^2} \Big|\tilde t_i\right)  \right) \nonumber \\
	 &= 
	 \beta +  \mathbb{E}_i\left( \frac{\sum^N_i \sum^{T_i}_t  \mathbb{E}\left(\tilde \varepsilon_{it}-\bar{\tilde \varepsilon}\right) \tilde t_{i}}{\sum^N_i \sum^{T_i}_t \tilde t_{i}^2} \Big|\tilde t_i  \right) \nonumber \\
	 &= 
	 \beta
\end{align}

%


\subsection{Discussion of the bias}

A crucial assumption for the fixed effects model to provide valid estimates of structural duration dependence in $y_{it}$ is that the within-error term $\tilde{\varepsilon}_{it}$ has zero conditional expectation given the within-time dimension $\tilde{t}_i$:

\begin{equation}
	\label{eq:condition}
	\mathbb{E}(\tilde\varepsilon_{it}|\tilde t_i) = 0
\end{equation}

In the within-specification, the regressor $\tilde{t}_i$ is obtained through a transformation that depends on the individual-specific number of observations $T_i$.  
For right-censored spells, $T_i$ equals the right-censoring duration $\bar{\tau}$ and is exogenous.  
For the remaining individuals, $T_i$ equals the completed duration of the spell, $\tau_i \leq \bar{\tau}$, and is endogenously determined.  
In this latter case, the last observation is directly related to attrition.  
Consequently, positive values of $\tilde{t}_i$ are mechanically more likely to be associated with exits from the observation sample.\footnote{
	For non-censored spells, the maximum within-individual time $\tilde{t}_i = T_i - \bar{t}_i$ coincides with an exit, while all other values $\tilde{t}_i < T_i - \bar{t}_i$ do not.  
	The maximal within-individual duration $\tilde{t}_i = T_i - \bar{t}_i$ is always positive but may vary depending on the total number of observations for individual $i$.
}

If the variable $y_{it}$ is also related to attrition, \textit{i.e.}, it is more likely to take certain values when individuals exit the observation sample, the within-error term $\tilde{\varepsilon}_{it}$ will mechanically correlate with the within-duration regressor $\tilde{t}_i$ through non-censored spells.  
In this case, the condition described in \autoref{eq:condition} does not hold, and the fixed effects approach produces biased estimates of structural duration dependence in $y_{it}$.

To illustrate this point, consider the following example.  
Assume that the dependent variable takes only two values, $y_{it} = 0,1$, and that it is directly related to the attrition mechanism, \textit{i.e.}, $y_{it}$ is an exit indicator.  
Each individual follows one of two sequences (1) $y_i = \{0,0,\ldots,0,1\}_{T_i}$ if $T_i = \tau_i \leq \bar{\tau}$ and an exit occurs, or (2) $y_i = \{0,0,\ldots,0,0\}_{T_i}$ if $T_i = \bar{\tau}$ and no exit occurs during the observation period.  
For simplicity, assume that $\beta = 0$, \textit{i.e.}, there is no structural duration dependence in the data-generating process of $y_{it}$.  
The linear specification of the model then reduces to
\begin{align}
	y_{it} = \alpha_i + \varepsilon_{it} 
	\hspace{20pt}
	\text{and}
	\hspace{20pt}
	\tilde y_{it} = \tilde  \varepsilon_{it},
\end{align} 
where the distribution of $\varepsilon_{it}$ is degenerate, with mass points at $1-\alpha_i$ and $-\alpha_i$ only.\footnote{
	Estimating such a model is relatively common, as it corresponds to a linear probability model specification.
}

For right-censored spells, the error terms are constant so that $\tilde{\varepsilon}_{it}$ and $\tilde{t}_i$ are uncorrelated: $\varepsilon_i = \{\varepsilon_{i1}, \varepsilon_{i2}, \ldots, \varepsilon_{i\bar{\tau}}\}$ = $\{-\alpha_i, -\alpha_i, \ldots, -\alpha_i\}$.  
For non-censored spells, the error terms vary within individuals: $\varepsilon_i = \{\varepsilon_{i1}, \varepsilon_{i2}, \ldots, \varepsilon_{i\tau_i}\} = \{-\alpha_i, -\alpha_i, \ldots, 1-\alpha_i\}$.  
In these spells, the last within-individual error $\tilde{\varepsilon}_{iT_i}$ is always positive and larger than all previous realizations, which are negative.  
This higher value of $\tilde{\varepsilon}_{iT_i}$ mechanically induces a positive correlation between $\tilde{\varepsilon}_{it}$ and $\tilde{t}_i$.  
Consequently, the fixed effects estimator produces a positively biased estimate of the structural duration dependence parameter, $\beta = 0$ (see \autoref{section:sim1} for a numerical simulation).

The above developments are summarized in the following proposition:

\vspace{10pt}

\noindent\fbox{%
	\parbox{\textwidth}{%
		\textbf{Proposition: The within-estimation duration bias}\\[5pt]
		Estimation of structural duration dependence relationships
		using the fixed effects estimator is unbiased as long as $\mathbb{E}(\tilde{\varepsilon}_{it} \mid \tilde{t}_i) = 0$.  
		If this condition is violated, \textit{i.e.}, if the expected value of the within-individual error term depends on the value of the within-individual time dimension,  
		the fixed effects estimator is subject to a \textit{within-estimation duration bias}.
	}
}

\vspace{10pt}

Whether the assumption required for unbiasedness of the fixed effects estimator holds depends on the nature of $y_{it}$.  
As illustrated above, the \textit{within-estimation duration bias} is likely to arise when the idiosyncratic term takes non-zero values in expectation at the time of exit from the observation sample.  
This typically occurs when the dependent variable serves as a proxy for sample attrition.



\section{Simulation: a job search process}

\label{section:sim2}

In this third section, we provide a concrete example of a setup in which the \textit{within-estimation duration bias} arises. 
After presenting the setup, we conduct Monte Carlo simulations showing that the extent of the bias can be sizeable for variables closely linked to the attrition mechanisms.


\subsection{Setup}

%
%

%
%
%

This simulation investigates the presence and magnitude of the \textit{within-estimation duration bias} in a job search environment similar to that of \cite{lalive2022ddjs}, where hiring follows a three-step process.  
First, unemployed individuals apply to vacancies posted by firms.  
Second, firms decide whether to call back applicants for job interviews.  
Third, interviews may result in job offers, which are automatically accepted by the applicants.

We assume access to longitudinal data on job search activity and outcomes.  
We observe all applications $A_{it}$ submitted by individual $i$ at multiple time periods of her unemployment spell $t \in \{1, \ldots, T_i\} \subset \mathbb{N}^+$.  
These periods typically correspond to weeks or months of elapsed unemployment.  
For each application $a$, we know whether it results sequentially in a callback $c_{ait}$ and a job offer $o_{ait}$.  
Job seekers are heterogeneous and characterized by a single time-invariant individual parameter $\alpha_i$.

This parameter influences both the application effort of job seekers ($A_{it}$) and their application-level probability of receiving a positive callback ($c_{ait}$).  
Similarly, elapsed unemployment duration directly affects the data-generating process of $A_{it}$ and $c_{ait}$, \textit{i.e.}, duration has a structural effect on both application effort and the callback rate.  
Job interviews are eventually converted into job offers through an exogenous process.  
Since job seekers always accept job offers, this last outcome characterizes the attrition mechanism and endogenously determines the number of periods over which individual $i$ is observed ($T_i = \tau_i \leq \bar{\tau}$ if the spell is non-censored, and $T_i = \bar{\tau}$ otherwise).


\subsection{Data generating process}

%
%

To simulate job-application data, we first generte individual heterogeneity parameters for a sample of \(N\) job seekers. 
Each individual \(i\) is assigned a single parameter \(\alpha_i\), drawn from the distribution \(\Gamma_{\alpha}(\cdot)\).
Given a maximum observable unemployment duration \(\bar{\tau}\), after which job seekers are no longer observed, 
we then generate application effort for each individual \(i\) and each period \(t \in \{1, \ldots, \bar{\tau}\} \subset \mathbb{N}^+\), according to the following data-generating process:
\begin{equation}
	\label{eq:sim2_A_it}
	A_{it} = \phi(\alpha_i, t , \xi_{it}; \beta) \in \mathbb{N}^+,
\end{equation}
where application effort depends on individuals' type $\alpha_i$,  elapsed unemployment duration $t$, operating through structural duration dependence, and any idiosyncratic error term $\xi_{it}$.

Based on the draws from \autoref{eq:sim2_A_it}, we generate $A_{it}$ individual applications for job seeker $i$ in unemployment period $t$. 
Each application results in either a positive callback or no callback, denoted by $c_{ait} \in \{0,1\}$. 
This binary outcome is governed by the following equation:

\begin{equation}
	c_{ait} = 
	\begin{cases}
		1 \hspace{10pt}\text{with probability} \hspace{10pt} \rho_c(\alpha_i,t ; \gamma)\\
		0 \hspace{10pt} \text{otherwise} 
	\end{cases}\,
\end{equation}

where the callback probability is influenced by both individual heterogeneity $\alpha_i$ and structural duration dependence.

By summing application-level callbacks at the individual?duration level, we obtain the number of job interviews in period $t$, which serves as a proxy for labor market matching:
\begin{equation}
	C_{it} = \sum_{a = 1}^{A_{it}} c_{ait} \in \mathbb{N}.
\end{equation}
At the end of the process, application-level callbacks are converted into job offers with exogenous probability $\psi$:
\begin{equation}
	o_{ait} = 
	\begin{cases}
		1 \hspace{10pt}\text{with probability} \hspace{10pt} \psi \hspace{10pt} \text{if}  \hspace{10pt} c_{ait} = 1\\
		0 \hspace{10pt}\text{otherwise} 
	\end{cases}\,
\end{equation}


%
The variable $o_{ait}$ therefore characterizes the attrition mechanism in the sample: individuals' job search activity and outcomes are observed only as long as they have not received a job offer.
In the empirically observed sample $\mathcal{S}_1$, job seekers are observed during periods $t \in \{1, \ldots, T_i\}$, with $T_i = \tau_i \leq \bar{\tau}$, after which they may exit the sample.  
In contrast, in the data-generating sample $\mathcal{S}_0$, individuals are observed throughout the entire potential observation period $t \in \{1, \ldots, \bar{\tau}\}$.


%
%

\subsection{Specification}

Following the above data generating process, we conduct Monte Carlo simulations with sample size $N = \DTLfetch{DynamicWriting2}{object}{N}{value}$ and $K = \DTLfetch{DynamicWriting2}{object}{K}{value}$ replications.
The right-censoring duration is set to $\bar \tau = 15$ months.
The functional forms we use are reported in \autoref{table:sim_2_calibration}. 

We calibrate the parameters following \cite{lalive2022ddjs} to replicate the average application behavior and callback rates during the first month of unemployment ($\overline{A}_1 = 10.5$ and $\overline{c}_1 = 0.05$), when no dynamic selection has yet occurred.
The parameters $\beta$, $\gamma$, and $\psi$ are set to values that reproduce the empirical patterns observed in the reference study.


%
%

\subsection{Results}

%
%

\begin{figure}[b!]
	\centering
	\caption{Simulation II, data generating processes and empirical duration profiles}
	\label{fig:simulation_2_GDP}
	\begin{subfigure}{0.495\textwidth}
		\caption{Applications $A_{it}$}
		\label{fig:1_1_DDJS_MC_saturated_A}
		\includegraphics[width=\textwidth]{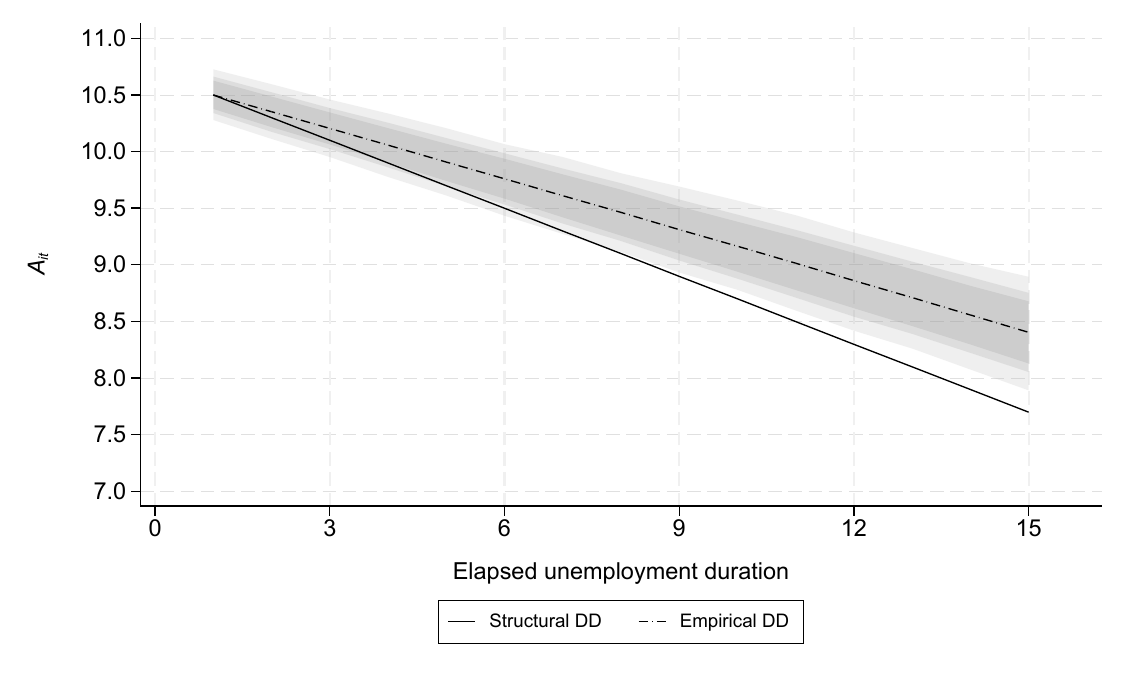}
	\end{subfigure}
	\begin{subfigure}{0.495\textwidth}
		\caption{Callback rate $c_{ait}$}
		\label{fig:1_1_DDJS_MC_saturated_c}
		\includegraphics[width=\textwidth]{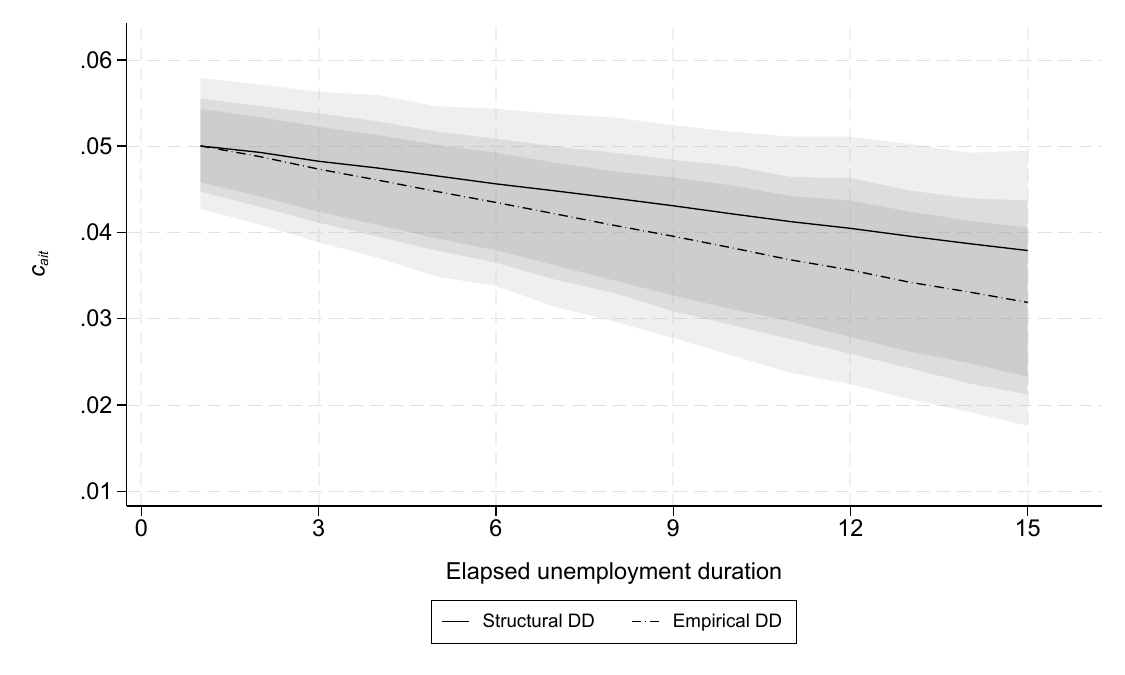}
	\end{subfigure}
	\begin{subfigure}{0.495\textwidth}
		\caption{Number of interviews $C_{it}$}
		\label{fig:1_1_DDJS_MC_saturated_C}
		\includegraphics[width=\textwidth]{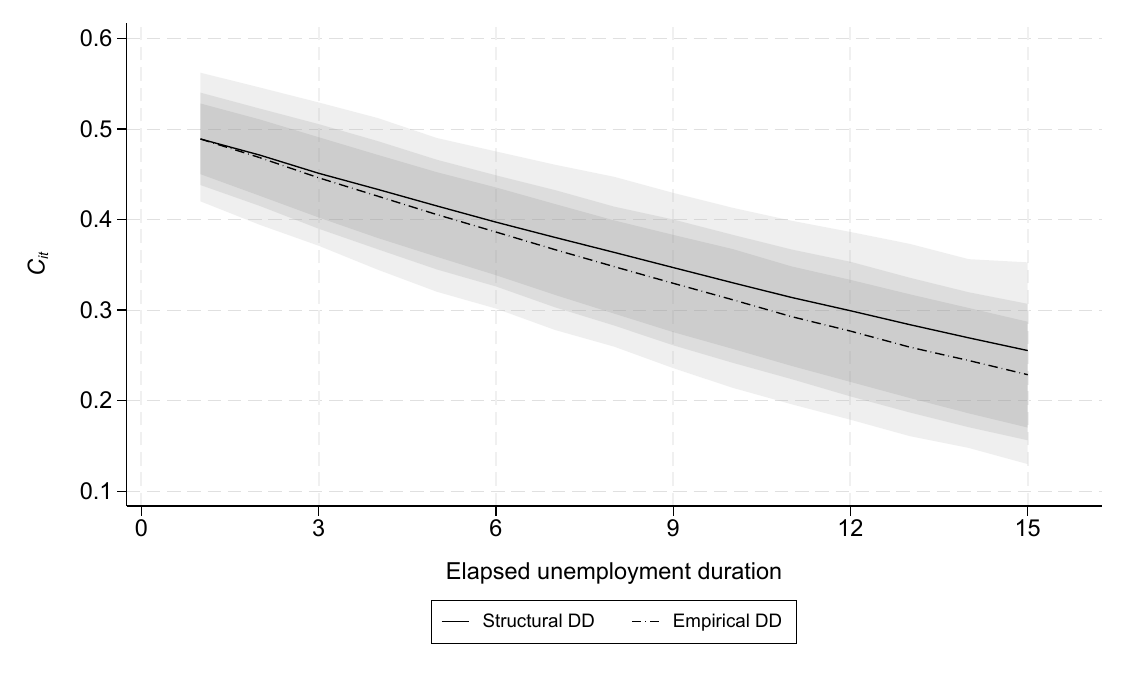}
	\end{subfigure}
	\ftnotefigure{
		Note: 
		This figure reports the structural duration dependence in $A_{it}$, $c_{ait}$ and $C_{it}$ (measured in sample $\mathcal{S}_0$), together with the patterns observed empirically (measured in sample $\mathcal{S}_1$).
		Percentiles of the distribution of the empirical duration profiles are also reported (1\textsuperscript{st}, 5\textsuperscript{th}, 10\textsuperscript{th}, 90\textsuperscript{th}, 95\textsuperscript{th} and 99\textsuperscript{th}). 
		Statistics are obtained from Monte Carlo simulations, with $K =  \DTLfetch{DynamicWriting2}{object}{K}{value}$ repetitions and sample size $N = \DTLfetch{DynamicWriting2}{object}{N}{value}$.
	}
\end{figure}

\autoref{fig:simulation_2_GDP} shows the structural duration dependence profiles from the data-generating process (solid lines) alongside the average duration patterns observed empirically (dashed lines) for the three outcomes $A_{it}$, $c_{ait}$, and $C_{it}$.
The figure also presents percentiles of the distributions of the empirical duration profiles.  
For all three variables, the structural duration dependence relationships differ substantially from their observed counterparts.  
For application effort, the empirical duration dependence tends to understate the negative structural effect of time on $A_{it}$.  
The opposite holds for the callback rate $c_{ait}$ and the number of callbacks $C_{it}$.
These findings reveal the presence of substantial dynamic selection: individuals with high $\alpha_i$, who apply more and face a lower chance of callbacks, tend to remain longer in the observed sample.  
This is confirmed by \autoref{fig:1_1_DDJS_MC_alpha_dynamic_selection} in the Appendix, which plots the average type $\alpha_i$ against elapsed unemployment duration for the samples with and without attrition ($\mathcal{S}_1$ and $\mathcal{S}_0$).

%
%

From the perspective of applied econometricians, we formulate the following linear models to approximate the data-generating processes:
\begin{align}
	\label{eq:lin_A_it}
	A_{it} &= \phi^A_{i} + f_{A}(t;\delta^{A}) + \varepsilon^A_{it} \\
	\label{eq:lin_c_ait}
	c_{ait} &= \phi^c_{i}+ f_{c}(t;\delta^{c})  + \varepsilon^c_{ait} \\
	\label{eq:lin_C_it}
	C_{it} &= \phi^C_{i} + f_{C}(t;\delta^{C})  + \varepsilon^C_{it},
\end{align}
where the functions $f(t;\delta)$ capture the structural effects of time to be estimated, $\phi$ represent heterogeneity components, and $\varepsilon$ are idiosyncratic error terms.  
We estimate these equations on the observed sample $\mathcal{S}_1$ using OLS with three different specifications:  
(1) a bivariate model with duration $t$ as the sole regressor,  
(2) a multivariate model including $\alpha_i$ as an additional control and  
(3) a model with individual fixed effects.\footnote{
	In the second specification, $\alpha_i$ is assumed to be observed and enters the equation linearly; in the third specification, $\alpha_i$ is accounted for through the inclusion of fixed effects.
}
The parametric functions $f(t;\delta)$ are specified either linearly, \textit{i.e.} $f(t;\delta) = \delta t$, or in a saturated form, \textit{i.e.} $f(t;\delta)$ is approximated by a set of dummies $\{\delta_t\}_{t=1}^{\bar{\tau}}$.

%
%

\begin{figure}[t!]
	\centering
	\caption{Simulation II, saturated duration specification $\{\delta_t\}_{1..\bar\tau}$}
	\label{fig:simulation_2_FE}
	\begin{subfigure}{0.495\textwidth}
		\caption{Applications $A_{it}$}
		\label{fig:1_1_DDJS_MC_saturated_A_it}
		\includegraphics[width=\textwidth]{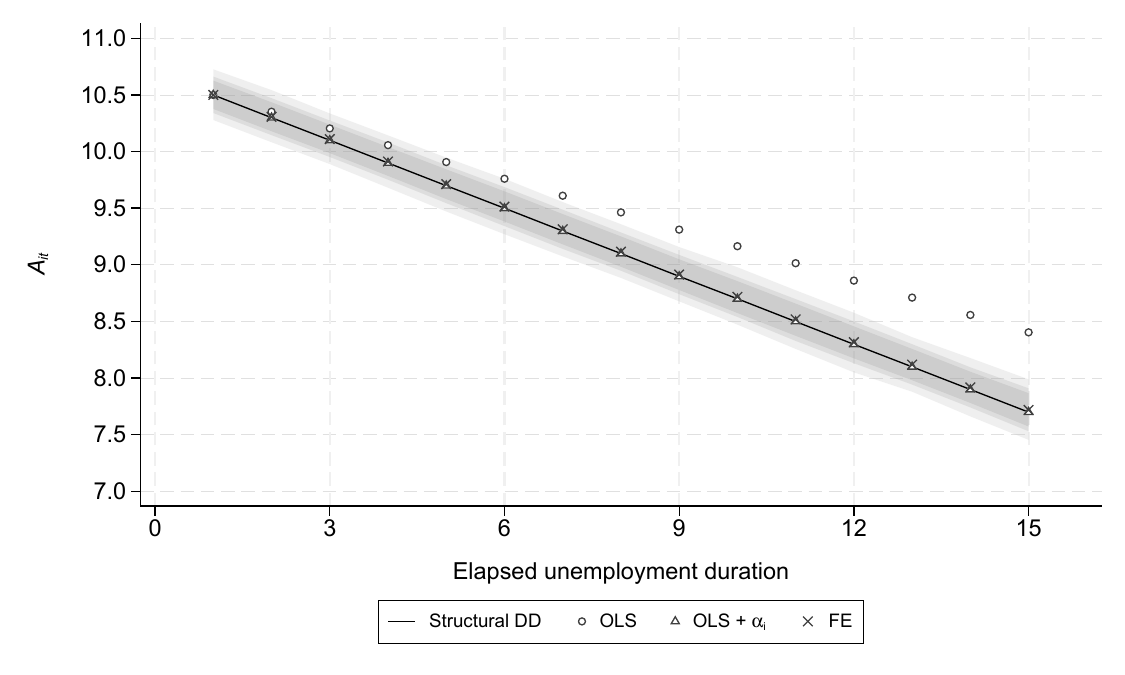}
	\end{subfigure}
	\begin{subfigure}{0.495\textwidth}
		\caption{Callback rate $c_{ait}$}
		\label{fig:1_1_DDJS_MC_saturated_c_ait}
		\includegraphics[width=\textwidth]{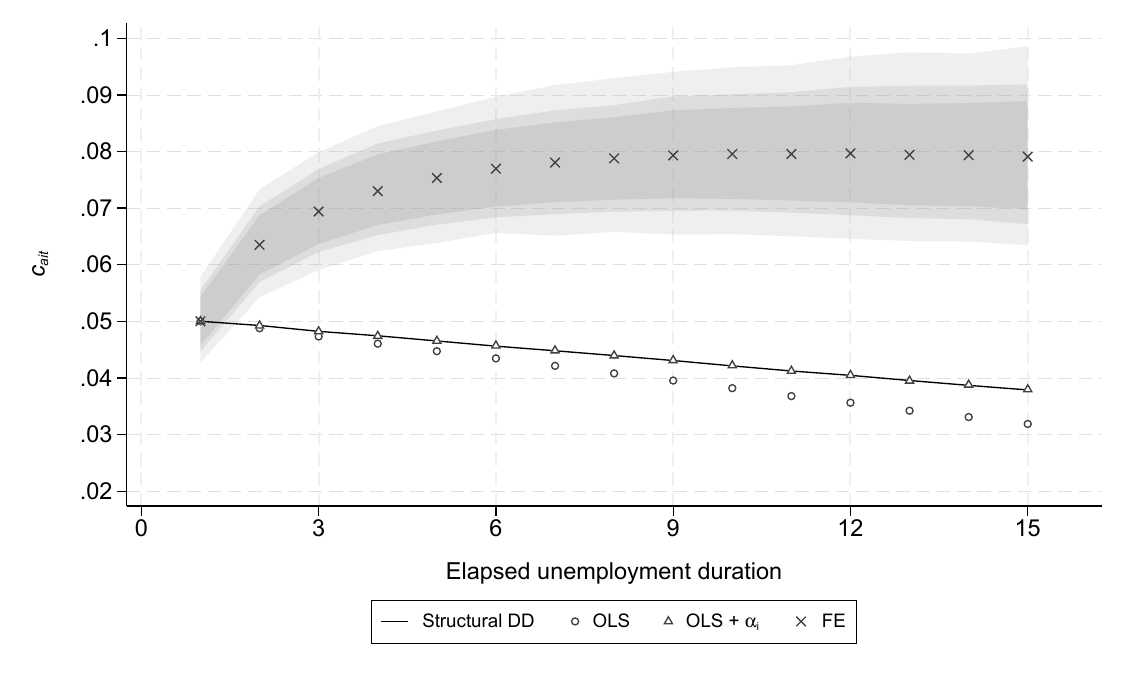}
	\end{subfigure}
	\begin{subfigure}{0.495\textwidth}
		\caption{Number of interviews $C_{it}$}
		\label{fig:1_1_DDJS_MC_saturated_C_it}
		\includegraphics[width=\textwidth]{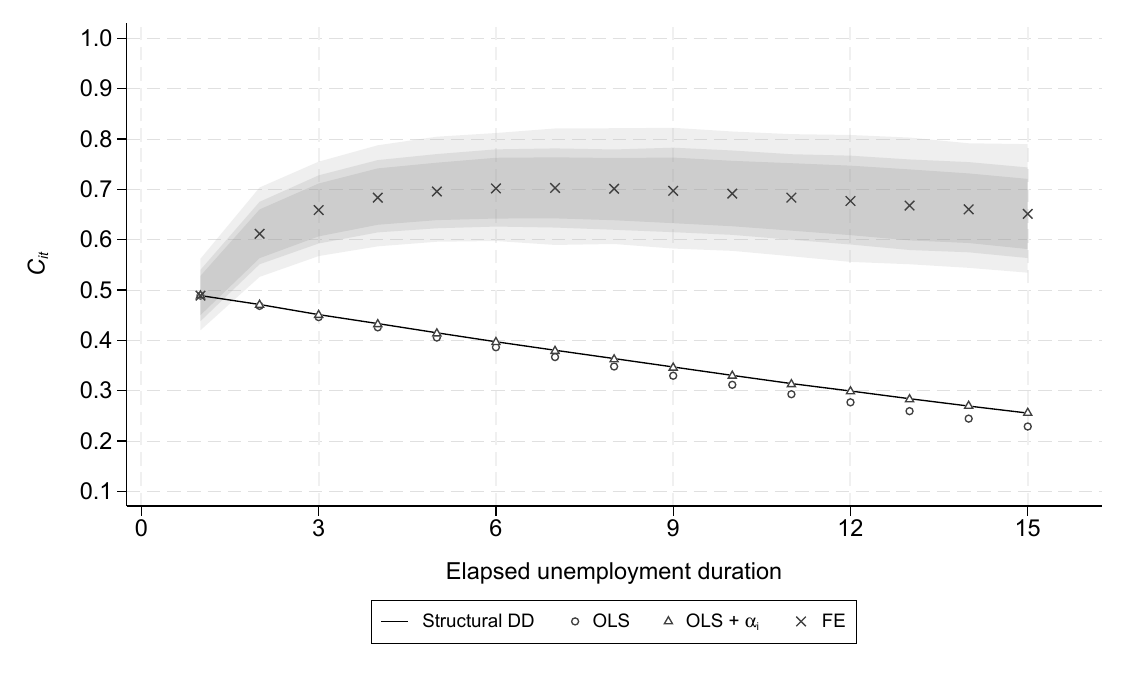}
	\end{subfigure}
	\ftnotefigure{
		Note: 
		This figure reports empirical estimates of structural duration dependence in $A_{it}$, $c_{ait}$ and $C_{it}$.
		The structural duration profiles to be estimated are depicted as solid lines.
		The parametric duration functions $f(t;\delta)$ are specified in a saturated manner, \textit{i.e.} $f(t;\delta) = \{\delta_t\}_{1..\bar\tau}$.
		Estimates are all based on the observation sample $\mathcal{S}_1$.
		Estimated duration profiles are obtained by regressing the outcomes on (1) time $t$ only (OLS), (2) time $t$ and $\alpha_i$ (OLS + $\alpha_i$) or (3) time $t$ and a set of individual fixed effects (FE).
		Distribution percentiles of the FE estimates are additionally reported (1\textsuperscript{st}, 5\textsuperscript{th}, 10\textsuperscript{th}, 90\textsuperscript{th}, 95\textsuperscript{th} and 99\textsuperscript{th}). 
		Statistics are obtained from Monte Carlo simulations, with $K =  \DTLfetch{DynamicWriting2}{object}{K}{value}$ repetitions and sample size $N = \DTLfetch{DynamicWriting2}{object}{N}{value}$.
	}
\end{figure}

Estimation results for the saturated specification are shown graphically in \autoref{fig:simulation_2_FE}, while complementary results for the linear specification are presented in \autoref{fig:linear_u_dur} in the Appendix.  
As with the empirical average duration profiles, the bivariate model including only duration as a regressor provides biased approximations of the structural duration relationships.  
In the absence of controls for individual heterogeneity, the estimates are upward-biased for $A_{it}$ and downward-biased for $c_{ait}$ and $C_{it}$.

The inclusion of $\alpha_i$ as a regressor largely addresses the dynamic selection issue: estimates from this second specification provide fairly accurate approximations of the structural duration profiles.\footnote{
	The linear model described in \autoref{eq:lin_A_it} and \autoref{eq:lin_c_ait} delivers an unbiased estimates of the structural duration profile in $A_{it}$ and $c_{ait}$, since both data-generating processes are linear in $\alpha_i$.  
	In contrast, the linear model \autoref{eq:lin_C_it} only approximate the structural duration dependence in $C_{it}$, as the data-generating process for this variable is non-linear in $\alpha_i$.
}
However, this approach relies on the observability of the heterogeneity component $\alpha_i$, which may be imperfect in practice.\footnote{
	In the Appendix, we report additional simulation results assuming that econometricians only have access to imperfect proxies for $\alpha_i$.  
	These proxies are generated as follows: $\tilde \alpha_i = \alpha_i + \eta_i$, where $\eta_i \sim \mathcal{N}(0, \sigma_\eta)$.
}

The fixed effects approach, which supposedly account for any form of time-invariant heterogeneity, produces mixed results.  
On the one hand, structural duration dependence in the number of job applications is well captured by the fixed effects model.  
On the other hand, within-estimates of the structural duration profiles are largely inaccurate for the callback rate and the number of callbacks.  

The substantial upward biases in the fixed effects estimates for $c_{ait}$ and $C_{it}$ arise from the close link between these variables and the attrition mechanism.
For these variables, the within transformation induces strong mechanical correlations between the within-time dimension and the within-error terms. As shown in \autoref{fig:1_2_xi} in the Appendix, the expected value of $\tilde \varepsilon_{it}$ for $c_{ait}$ and $C_{it}$ is largely positive for positive values of $\tilde t$ (or equivalently, for $t = T_i$) in the subsample of non-censored spells.
In contrast, $A_{it}$ is only weakly related to attrition, which explains the absence of a noticeable bias in the estimated structural duration profile for this variable.\footnote{
	Qualitatively, the error term of $A_{it}$ is also systematically higher in periods when attrition occurs: job seekers idiosyncratically apply more when they exit the sample.  
	However, this pattern is quantitatively negligible, explaining the virtually absent bias in the structural duration profile of $A_{it}$ estimated by the fixed effects model.
}
The \textit{within-estimation duration bias} may nonetheless arise for application effort if job seekers increase their applications sharply at the time they exit unemployment, for idiosyncratic reasons.\footnote{
	This can occur, for example, if job search effort rises idiosyncratically due to the exhaustion of unemployment benefits (\citealp{marinescu2021unemployment, dellavigna2022evidence}).  
	If this coincides with a higher attrition rate, the fixed effects duration estimates for job applications may also be biased.
}

The above results highlight the importance of carefully examining the nature of the dependent variables before applying the fixed effects estimator to estimate structural duration dependence relationships.  
If the dependent variable is not directly linked to the attrition mechanism, the fixed effects approach adequately accounts for unobserved heterogeneity and provides unbiased estimates of the structural effect of time.  
In contrast, when the variable of interest is directly related to attrition, the within-estimation procedure can create spurious correlations between the dependent variable and the time regressor, leading to substantial bias in the estimated structural duration profile.  
In such cases, linear models should instead be specified as multivariate models that include appropriate controls for individual heterogeneity.



\section{Conclusion}

\label{section:conclusion}

Recent work in applied labor economics has made use of innovative longitudinal job search data to study duration dependence phenomena.  
Access to such data ostensibly facilitates the estimation of the structural effect of time, as it allows researchers to control for all time-invariant sources of heterogeneity using fixed effects models. 
However, related studies have often overlooked the specific nature of the time regressor and the fact that it lies at the core of the within-transformation applied by these models.

In this paper, we derive the conditions under which the linear fixed effects model provides valid estimates of structural duration profiles.  
We show that fixed effects estimates can be biased when the value of the within-error term depends on the within-time regressor.  
This bias particularly arises when the independent variable is related to the attrition mechanism affecting the sample.  
Using Monte Carlo simulations, we show that the magnitude of this \textit{within-estimation duration bias} can be extremely large.  
Considering a stylized job search environment, we find that structural duration profiles obtained from fixed-effects models are substantially biased for outcomes directly affected by sample attrition, such as callbacks or callback rates.
In contrast, unbiased estimates are obtained for variables whose within-error term does not correlate with the within-time dimension, such as applications.

From an applied perspective, this paper suggests that fixed effects models are not a ``silver bullet'' for estimating structural duration dependence in the presence of individual heterogeneity and dynamic selection.  
These models should not be applied blindly when longitudinal data are available; their use should instead be guided by theory and by researchers? knowledge of the nature of the dependent variable whose duration profile is under study.  
If the outcome variable is not directly related to the attrition mechanism, the fixed effects approach can provide unbiased estimates of the structural effect of time.  
Otherwise, the within-estimator is likely to be inaccurate, and specifications that include controls for heterogeneity should be preferred when approximating structural duration profiles.



\nocite{*}
\bibliographystyle{apalike}
\bibliography{./Literature/DDFE_literature}

 

\clearpage
\appendix

\renewcommand{\thesection}{\Alph{section}}
\newcommand{\thesectionappendix}{\Alph{section}}
\newcommand{\theequationappendix}{}

\renewcommand{\thesubsection}{\Alph{section}.\arabic{subsection}}

\counterwithin{figure}{section}
\renewcommand\thefigure{\thesectionappendix\arabic{figure}}  

\counterwithin{table}{section}
\renewcommand\thetable{\thesectionappendix\arabic{table}} 

\numberwithin{equation}{section}
\renewcommand{\theequationappendix}{\arabic{equation}}



\clearpage

\section{The within-estimation duration bias}

\label{section:theory_appendix}

Based on the within-data generating process and the OLS estimator definition, we write the fixed effects estimator of $\beta$ as
\begin{align*}
	\hat\beta_{FE}&= \frac{\sum^N_i \sum^{T_i}_t (\tilde y_{it} -  \bar{\tilde y}) (\tilde t_{i} -  \bar{\tilde t})}{\sum^N_i \sum^{T_i}_t (\tilde t_{i} -  \bar{\tilde t})^2} \nonumber
	= \frac{\sum^N_i \sum^{T_i}_t \left[\beta (\tilde t_{i} -  \bar{\tilde t}) +  (\tilde \varepsilon_{it} -  \bar{\tilde \varepsilon}) \right] (\tilde t_{i} -  \bar{\tilde t})}{\sum^N_i \sum^{T_i}_t (\tilde t_{i} -  \bar{\tilde t})^2} \nonumber \\[5pt]
	&= \beta + 	\frac{\sum^N_i \sum^{T_i}_t  (\tilde \varepsilon_{it} -  \bar{\tilde \varepsilon}) (\tilde t_{i} -  \bar{\tilde t})}{\sum^N_i \sum^{T_i}_t (\tilde t_{i} -  \bar{\tilde t})^2} 
	= \beta + 	\frac{\sum^N_i \sum^{T_i}_t  (\tilde \varepsilon_{it} -  \bar{\tilde \varepsilon}) \tilde t_{i}}{\sum^N_i \sum^{T_i}_t \tilde t_i^2}
\end{align*}

The last equality come from	$\bar{\tilde t} 
= \frac{\sum_{i}^{N}(\frac{1}{T_i}\sum_{t}^{T_i} \tilde t_i)}{\sum_{i}^N T_i} 
$, where each individual-specific sum can be rewritten as:
\begin{align*}
	\sum^{T_i}_{t=1} \tilde t_{i} 
	= \sum_{t=1}^{T_i} \left(t-\frac{T_i+1}{2}\right) 
	= -\frac{T_i(T_i+1)}{2} + \sum_{t=1}^{T_i} t
\end{align*}

It can be proven by induction that $\sum_{t=1}^{T_i} t = \frac{T_i(T_i+1)}{2}$.
First, this equation holds for  $T_i = 1$: $\frac{1}{1}\sum^1_{t=1}t = 1 = \frac{1+1}{2}$.
Second, the equation holds for any $T_i + 1$, assuming that the relationship $\frac{1}{T_i}\sum^{T_i}_{i=1} t  = \frac{T_{i}+1}{2}$, or similarly $\sum^{T_i}_{i=1} t = \frac{T_i(T_i+1)}{2}$, holds for $T_i$:
\begin{align*}
	\frac{1}{T_i+1}\sum^{T_i+1}_{t=1}	 t 
	&= \frac{1}{T_i+1}\left(\sum^{T_i}_{t=1} t + (T_i + 1)\right)   
	= \frac{\left(\frac{T_i(T_i+1)}{2} + (T_i + 1)\right) }{T_i+1}
	= \frac{T_i}{2} + 1 
	= \frac{(T_i + 1) + 1}{2}
\end{align*}
Consequently, 
\begin{align*} 
	T_i\bar{\tilde t}_{i} 
	= -\frac{T_i(T_i+1)}{2} + \sum_{t=1}^{T_i} t
	= -\frac{T_i(T_i+1)}{2}  + \frac{T_i(T_i+1)}{2}  = 0
\end{align*}

Taking the expectation of $\hat\beta_{FE}$, we get:
\begin{align*}
	\mathbb{E}\left(\hat \beta_{FE} \right) 
	&= \mathbb{E}\left(\beta + 	\frac{\sum^N_i \sum^{T_i}_t  (\tilde \varepsilon_{it} -  \bar{\tilde \varepsilon}) \tilde t_{i}}{\sum^N_i \sum^{T_i}_t \tilde t_{i}^2} \right)  = \beta + \mathbb{E}\left(\frac{\sum^N_i \sum^{T_i}_t  (\tilde \varepsilon_{it} -  \bar{\tilde \varepsilon}) \tilde t_{i}}{\sum^N_i \sum^{T_i}_t \tilde t_{i}^2} \right)
\end{align*}

Assuming $\mathbb{E}(\tilde\varepsilon_{it}|\tilde t_i) = 0$, we get an unbiased estimator of $\beta$:
\begin{align*}
	\mathbb{E}\left(\frac{\sum^N_i \sum^{T_i}_t  \tilde \varepsilon_{it} \tilde t_{i}}{\sum^N_i \sum^{T_i}_t \tilde t_{i}^2} \right) 
	&= \mathbb{E}_i\left( \mathbb{E}\left(\frac{\sum^N_i \sum^{T_i}_t  \tilde \varepsilon_{it} \tilde t_{i}}{\sum^N_i \sum^{T_i}_t \tilde t_{i}^2} \Big|\tilde t_i\right)  \right) 
	= \mathbb{E}_i\left( \frac{\sum^N_i \sum^{T_i}_t \mathbb{E}\left( \tilde \varepsilon_{it}  |\tilde t_i \right) \tilde t_{i}}{\sum^N_i \sum^{T_i}_t \tilde t_{i}^2}\right)  = 0,
\end{align*}
and similarly for $\mathbb{E}\left(\frac{\sum^N_i \sum^{T_i}_t  \bar{\tilde \varepsilon} \ \tilde t_{i}}{\sum^N_i \sum^{T_i}_t \tilde t_{i}^2} \right) $. \\[0pt]

As long as the above condition is met, the fixed effects estimator of the structural duration dependence parameter is unbiased: $\mathbb{E}\left(\hat \beta_{FE} \right)  = \beta$.


\clearpage


\section{Simulation I: a binary exit indicator}

\label{section:sim1}


\subsection{Setup}

This simulation examines the case of an exit indicator, as described in \autoref{section:bias}.
We consider a population of individuals who differ with respect to heterogeneity parameter $\alpha_i$ and who face a risk of leaving the population of interest with individual-specific and time-varying hazard $\lambda_{it}$.

We assume we have access to individual longitudinal data, where each sampled individual $i$ is observed a discrete number of time, at duration $t \in \{1,...,T_i\} \subset \mathbb{N}^+$.
The individual-specific number of observations is either the completed duration of the spell $T_i = \tau_i$ if $\tau_i \leq \bar \tau$, or the right-censoring duration $T_i = \bar \tau$ if $\tau_i > \bar \tau$.
Each individual is characterized by the series $y_{i} = \{0,0,...,0,1\}_{\tau_i}$ if the exit occurs before or at the right-censoring duration, or $y_{i} = \{0,0,...,0,0\}_{\bar\tau}$ otherwise.


\subsection{Data generating process}

\renewcommand{\arraystretch}{1.5}
\begin{table}[b!]
	\centering
	\caption{
		Simulation I,
		specification and calibration
	}
	\label{table:sim_1_calibration}
	\footnotesize
	\begin{tabularx}{\textwidth}{| X | L{5.0cm} | L{5.0cm} |} 
		\cline{2-3}
		\multicolumn{1}{c|}{} 
		& \textbf{Specification}
		& \textbf{Calibration} \\
		\hline
		\textbf{DGP}
		& 
		$\lambda_{it} = \frac{1}{1+\exp[\alpha_i + \gamma (t-1)]}$
		& $\gamma \in \{ 0.05, 0.00, -0.02 \}$ \\[5pt] \hline
		\textbf{Heterogeneity}
		& $\alpha_i = \alpha + \nu_i$ 
		& $\alpha = 2$
		\\ 
		& $\nu_i \sim \mathcal{N}(0,\sigma^2_\nu)$ 
		& $\sigma^2_\nu = 0.5$
		\\ \hline 
	\end{tabularx}
	\vspace{10pt}
	\ftnotetable{
		Note: 
		This table reports the specification and calibration used in Simulation I.
	}
\end{table}
\renewcommand{\arraystretch}{1.0}

We consider a sample made of $N$ individuals. Heterogeneity parameters $\alpha_i$ are drawn from a distribution $\Gamma_\alpha(\cdot)$.
For each agent, we generate $\bar \tau$ observations covering each period of the potential observation window, $t \in \{1,...,\bar \tau \} \subset \mathbb{N}^+$.
Exits are generated from a discrete-time survival model, where the hazard $\lambda_{it}$ and survival probability $\Lambda_{it}$ functions are defined as follows:
\begin{align}
	\label{eq:hazard}
	\lambda_{it} &= \mathbb{P}(\tau_i = t | \tau_i \geq t) = \lambda(\alpha_i,t;\gamma) \\
	\Lambda_{it} &= (1-\lambda_{i1} )(1-\lambda_{i2} )...(1-\lambda_{it-1}) = \prod^{t-1}_{k=1}(1-\lambda_{ik}).
\end{align}
The above equations show that the hazard and survival probability functions are heterogeneous across agents through $\alpha_i$. 
Duration also structurally affects the exit rate: if $\lambda'(t) < 0$, the individual-specific hazard decreases over time; it increases if $\lambda'(t) > 0$, and remains constant if $\lambda'(t) = 0$. 
The exact exit time is obtained using the \textit{Inverse Transform Sampling} method applied to the survival function $\Lambda_{it}$. 
This yields a unique time $\tau_i$ at which individual $i$ exits the observed sample $\mathcal{S}_1$.  
In contrast, in the data-generating sample $\mathcal{S}_0$, individuals are observed over the entire potential observation period $t \in \{1,\dots,\bar\tau\}$.

%
%
%
%
%

Based on the above setup, we conduct Monte Carlo simulations with sample size 
$N = \DTLfetch{DynamicWriting1}{object}{N}{value}$ and $K = \DTLfetch{DynamicWriting1}{object}{K}{value}$ replications.
The maximum observation period is set to $\bar{\tau} = 15$.
The functional forms and calibration choices used for the data-generating process are described in \autoref{table:sim_1_calibration}.


\subsection{Results}

\begin{figure}[b!]
	\centering
	\caption{Simulation I, data generating process and empirical duration profiles}
	\label{fig:1_2_MC_DGP}
	\begin{subfigure}{0.495\textwidth}
		\caption{$\gamma = 0.05$}
		\label{fig:1_2_MC_DGP_5}
		\includegraphics[width=\textwidth]{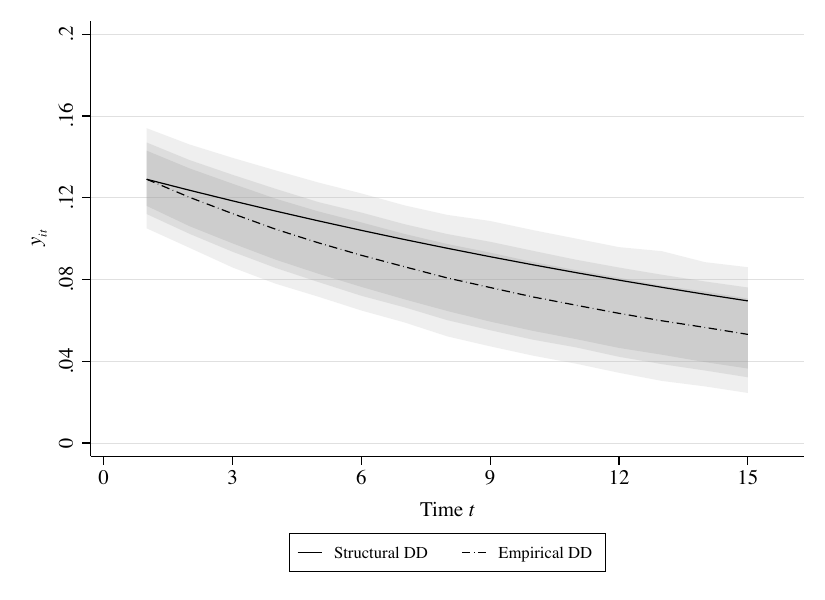}
	\end{subfigure}
	\begin{subfigure}{0.495\textwidth}
		\caption{$\gamma = 0.00$}
		\label{fig:1_2_MC_DGP_0}
		\includegraphics[width=\textwidth]{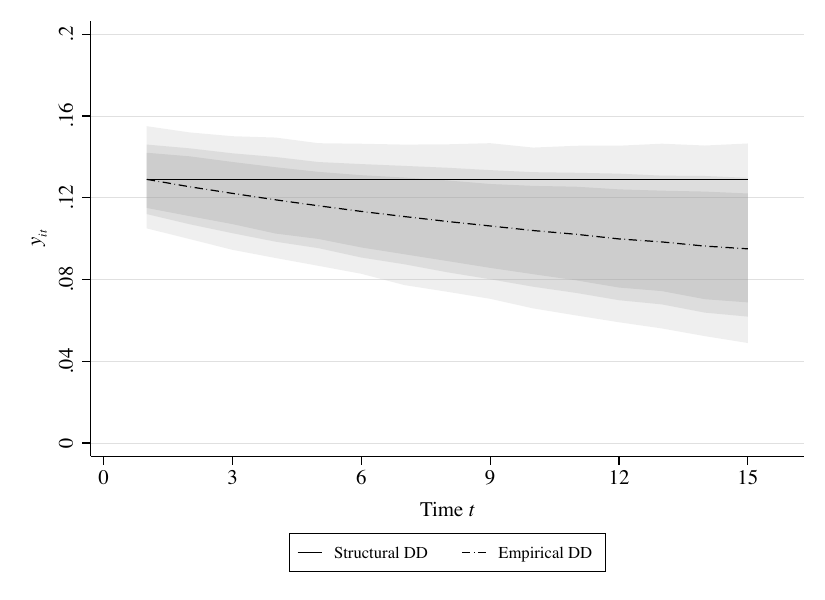}
	\end{subfigure}
	\begin{subfigure}{0.495\textwidth}
		\caption{$\gamma = -0.02$}
		\label{fig:1_2_MC_DGP_}
		\includegraphics[width=\textwidth]{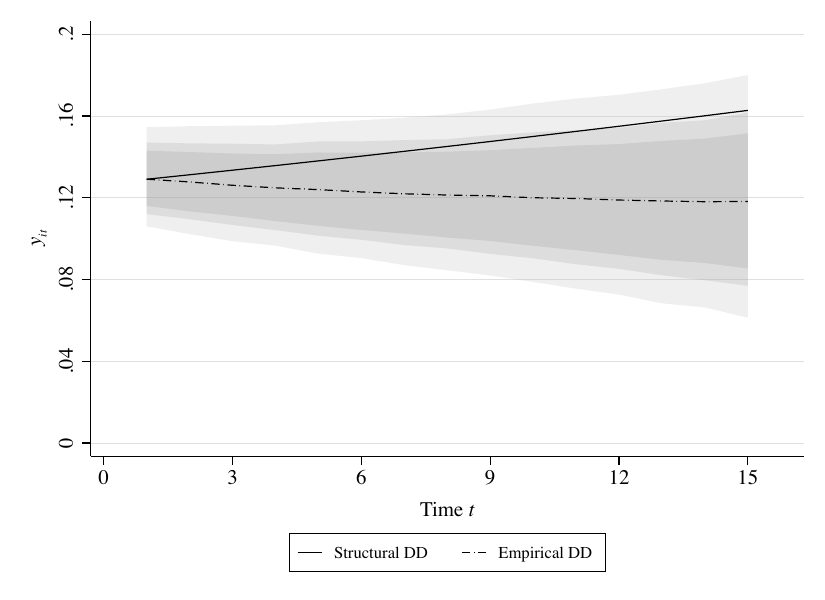}
	\end{subfigure}
	\ftnotefigure{
		Note: 
		This figure reports the structural duration dependence in the exit rate (measured by the average $\lambda_{it}$ per duration $t$, in sample $\mathcal{S}_0$), together with the average empirical hazard rate (measured by the average $y_{it}$ per duration $t$, in sample $\mathcal{S}_1$).
		Three different values of the structural duration parameter are considered: $\gamma = 0.05, 0.00$ and $-0.02$.
		Percentiles of the distribution of the empirical duration profiles are also reported (1\textsuperscript{st}, 5\textsuperscript{th}, 10\textsuperscript{th}, 90\textsuperscript{th}, 95\textsuperscript{th} and 99\textsuperscript{th}). 
		Statistics are obtained from Monte Carlo simulations, with $K =  \DTLfetch{DynamicWriting1}{object}{K}{value}$ repetitions and sample size $N = \DTLfetch{DynamicWriting1}{object}{N}{value}$.
	}
\end{figure}

\begin{figure}[b!]
	\centering
	\caption{Simulation I, dynamic selection}
	\label{fig:1_2_MC_dynamic_selection}
	\begin{subfigure}{0.495\textwidth}
		\caption{$\gamma = 0.05$}
		\label{fig:1_2_MC_dynamic_selection_5}
		\includegraphics[width=\textwidth]{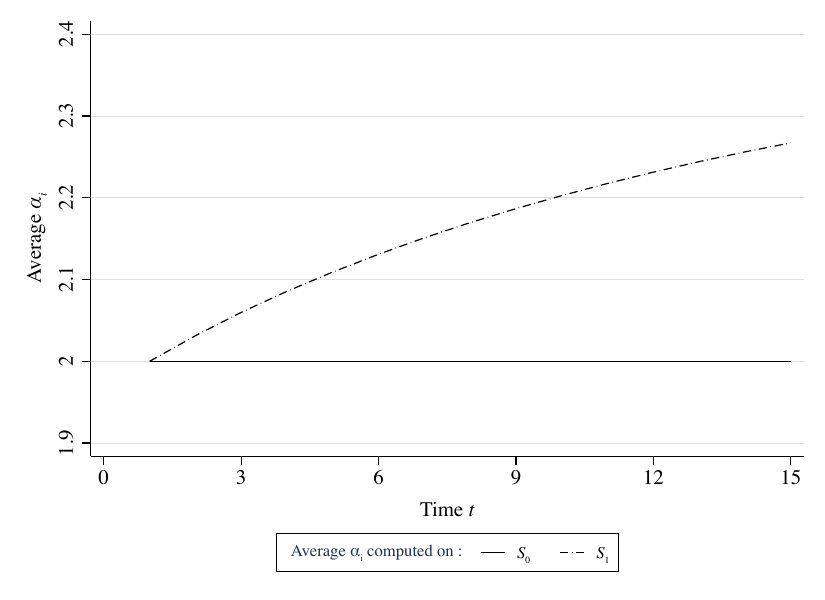}
	\end{subfigure}
	\begin{subfigure}{0.495\textwidth}
		\caption{$\gamma = 0.00$}
		\label{fig:1_2_MC_dynamic_selection_0}
		\includegraphics[width=\textwidth]{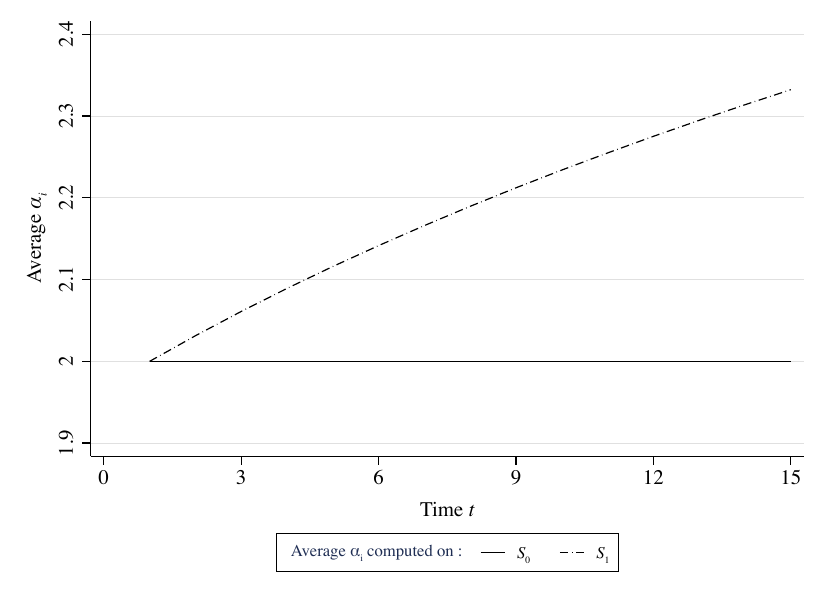}
	\end{subfigure}
	\begin{subfigure}{0.495\textwidth}
		\caption{$\gamma = -0.02$}
		\label{fig:1_2_MC_dynamic_selection_-2}
		\includegraphics[width=\textwidth]{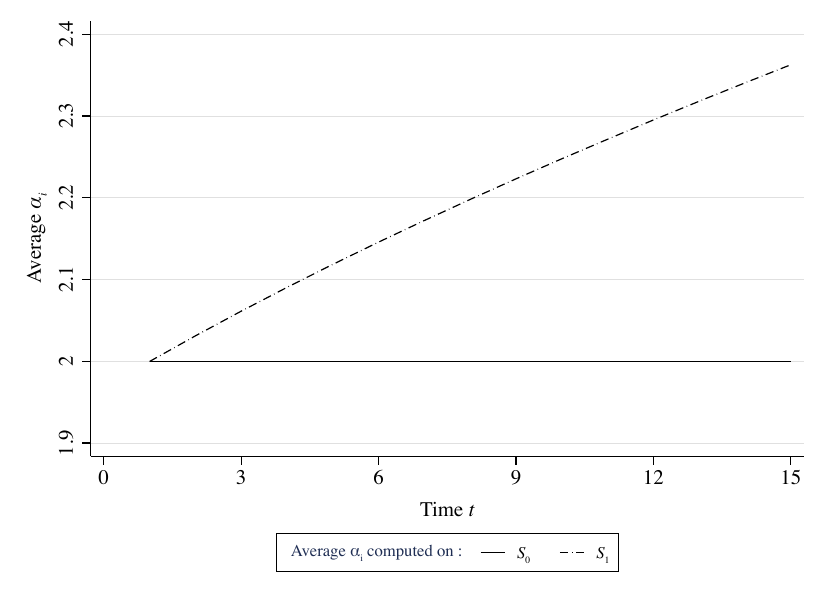}
	\end{subfigure}
	\ftnotefigure{
		Note: This figure presents evidence of dynamic selection for Simulation I.
		It reports the average $\alpha_i$ per duration $t$ in the data-generating sample $\mathcal{S}_{0}$ and in the observation sample $\mathcal{S}_{1}$, for three different values of the structural duration dependence parameter $\gamma = 0.05, 0.00$ and $-0.02$.
		Statistics are obtained from Monte Carlo simulations, with $K =  \DTLfetch{DynamicWriting1}{object}{K}{value}$ repetitions and sample size $N = \DTLfetch{DynamicWriting1}{object}{N}{value}$.
	}
\end{figure}

In \autoref{fig:1_2_MC_DGP}, we report the average structural duration-dependence profile of the exit rate (solid lines), together with the average empirical hazard rate (dashed lines), for different values of the structural duration parameter: $\gamma = 0.05$, $0.00$, and $-0.02$.\footnote{
	The average structural duration profile is obtained by taking the average of $\lambda_{it}$ for each period $t$, based on the sample without attrition ($\mathcal{S}_0$).
	The average empirical duration profile is obtained by taking the average of the exit indicator $y_{it}$ for each period $t$, based on the sample with attrition ($\mathcal{S}_1$).
}
The figure also reports the percentiles of the empirical duration-profile distributions.
Empirical duration profiles are systematically more negative than their structural counterparts across all panels.
This pattern is driven by dynamic selection:
as shown in \autoref{fig:1_2_MC_dynamic_selection}, individuals with high $\alpha_i$ are more likely to remain in the observation sample $\mathcal{S}_1$ for longer periods, as they face lower hazards at all durations.

\begin{figure}[t!]
	\centering
	\caption{Simulation I, saturated duration specification $\{\delta_t\}_{1..\bar\tau}$}
	\label{fig:1_2_MC_OLS_FE}
	\begin{subfigure}{0.495\textwidth}
		\caption{$\gamma = 0.05$}
		\label{fig:1_2_MC_OLS_FE_5}
		\includegraphics[width=\textwidth]{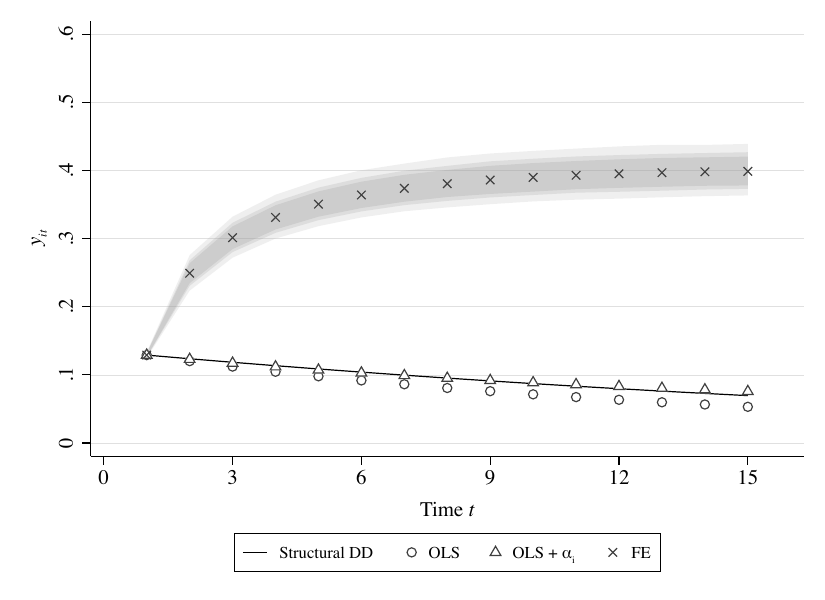}
	\end{subfigure}
	\begin{subfigure}{0.495\textwidth}
		\caption{$\gamma = 0.00$}
		\label{fig:1_2_MC_OLS_FE_0}
		\includegraphics[width=\textwidth]{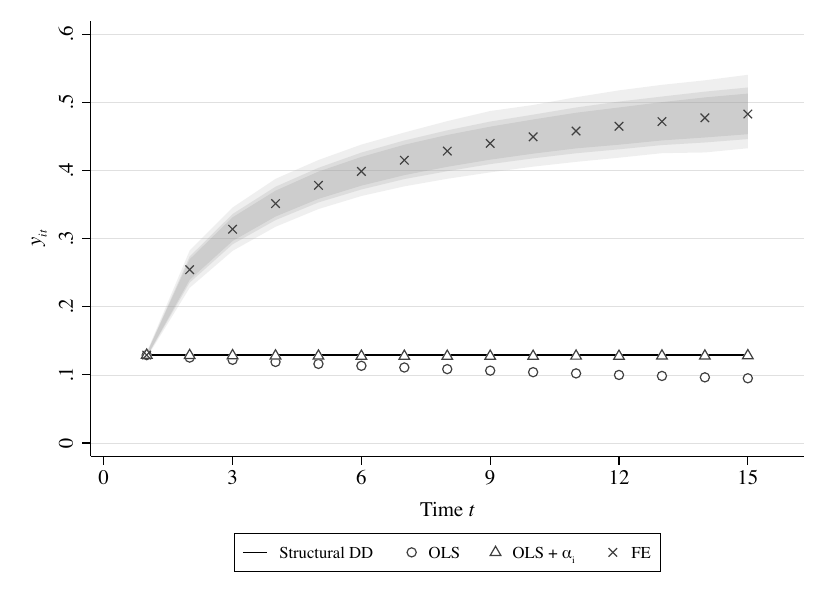}
	\end{subfigure}
	\begin{subfigure}{0.495\textwidth}
		\caption{$\gamma = -0.02$}
		\label{fig:1_2_MC_OLS_FE_-2}
		\includegraphics[width=\textwidth]{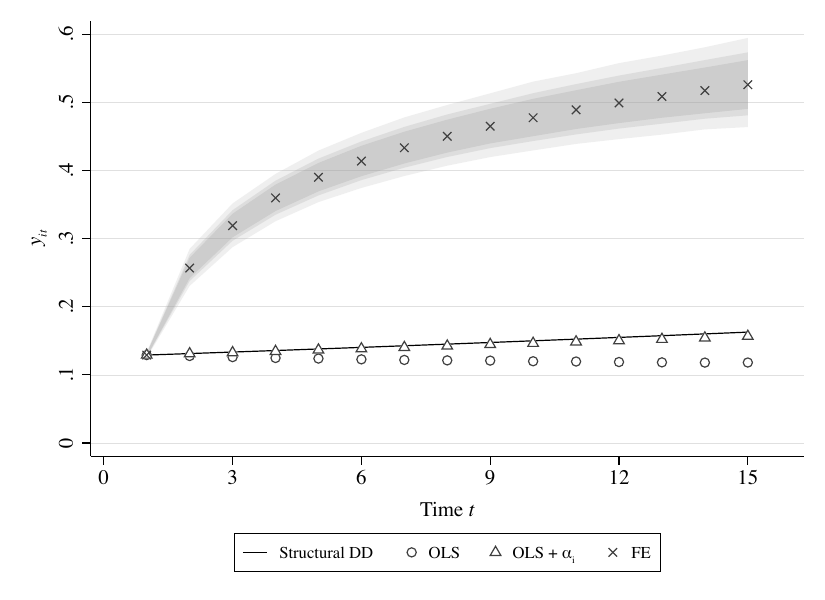}
	\end{subfigure}
	\ftnotefigure{
		Note: 
		This figure reports empirical estimates of structural duration dependence in the exit rate.
		Three different values of the structural duration dependence parameter are considered, $\gamma = 0.05, 0.00$ and $-0.02$.
		The structural duration profiles to be estimated are depicted as solid lines.
		Estimates are all based on the observation sample $\mathcal{S}_1$
		and are obtained by regressing $y_{it}$ on (1) time $t$ only (OLS), (2) time $t$ and $\alpha_i$ (OLS + $\alpha_i$) or (3) time $t$ and a set of individual fixed effects (FE).
		Distribution percentiles of the FE estimates are additionally reported (1\textsuperscript{st}, 5\textsuperscript{th}, 10\textsuperscript{th}, 90\textsuperscript{th}, 95\textsuperscript{th} and 99\textsuperscript{th}). 
		Statistics are obtained from Monte Carlo simulations, with $K =  \DTLfetch{DynamicWriting1}{object}{K}{value}$ repetitions and sample size $N = \DTLfetch{DynamicWriting1}{object}{N}{value}$.
	}
\end{figure}

Taking the perspective of applied econometricians, we seek to estimate structural duration dependence in the dependent variable using the observations contained in sample $\mathcal{S}_1$. 
To approximate the data-generating process, we specify the following linear model:
\begin{align}
	\label{eq:lin_y_it}
	y_{it} &= \phi_i + f(t;\delta) + \varepsilon_{it},
\end{align}
where $f(t;\delta)$ captures the structural effect of time on $y_{it}$, $\phi_i$ denotes individual-specific heterogeneity, and $\varepsilon_{it}$ is an idiosyncratic error term. 
We estimate several specifications of this model using OLS: 
(1) a bivariate model in which outcomes are regressed solely on $t$; 
(2) a multivariate model that additionally includes $\alpha_i$ to control for individual heterogeneity; and 
(3) a model that accounts for heterogeneity through individual fixed effects.
The parametric duration function $f(t;\delta)$ is specified in a saturated manner, \textit{i.e.}, $f(t;\delta)$ is approximated by a full set of time dummies $\{\delta_t\}_{t=1}^{\bar\tau}$.

\begin{figure}[h!]
\centering
\caption{Simulation I, correlation between $\varepsilon$, $\tilde t_{i}$ and $t-\tau_i$}
\label{fig:1_1_xi}
\begin{subfigure}{\textwidth}
	\caption{$\gamma = 5$}
	\label{fig:1_1_xi_5}
	\includegraphics[width=0.495\textwidth]{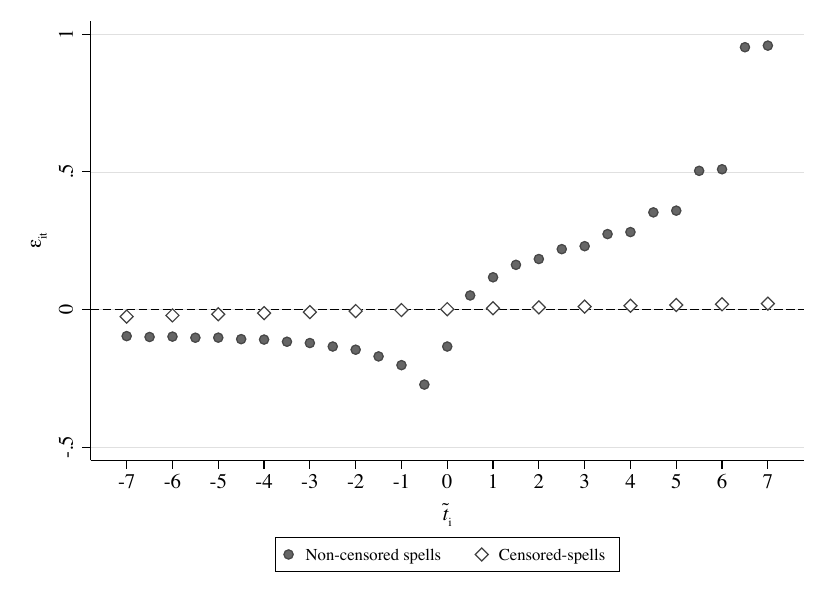}
	\includegraphics[width=0.495\textwidth]{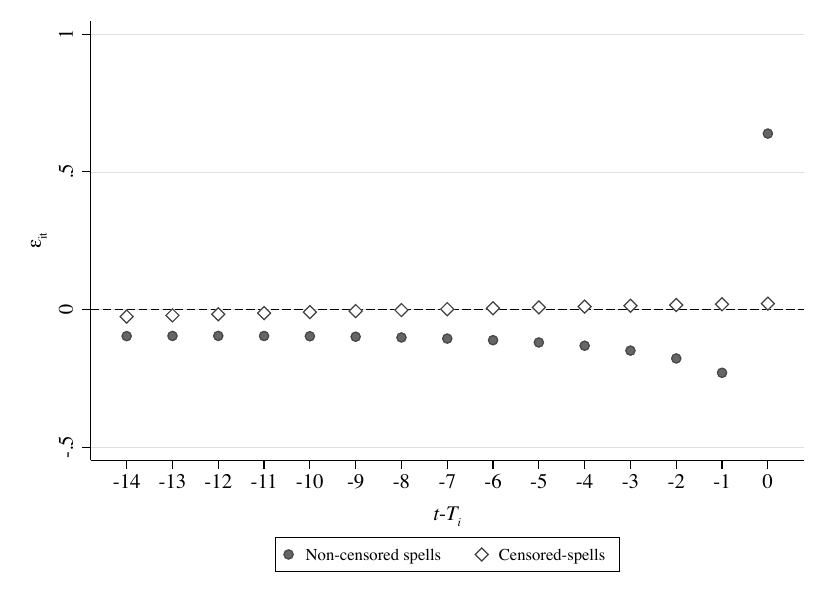}
\end{subfigure}
\begin{subfigure}{\textwidth}
	\caption{$\gamma = 0$}
	\label{fig:1_1_xi_0}
	\includegraphics[width=0.495\textwidth]{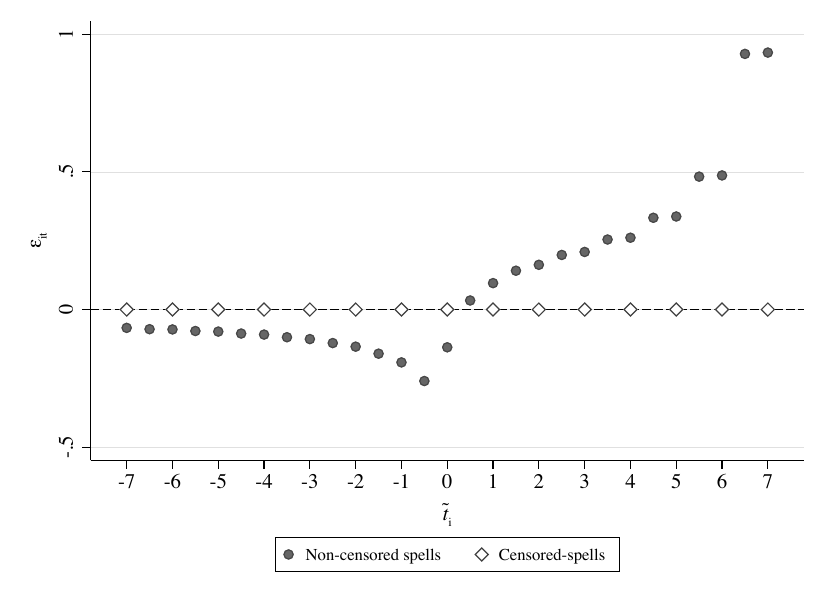}
	\includegraphics[width=0.495\textwidth]{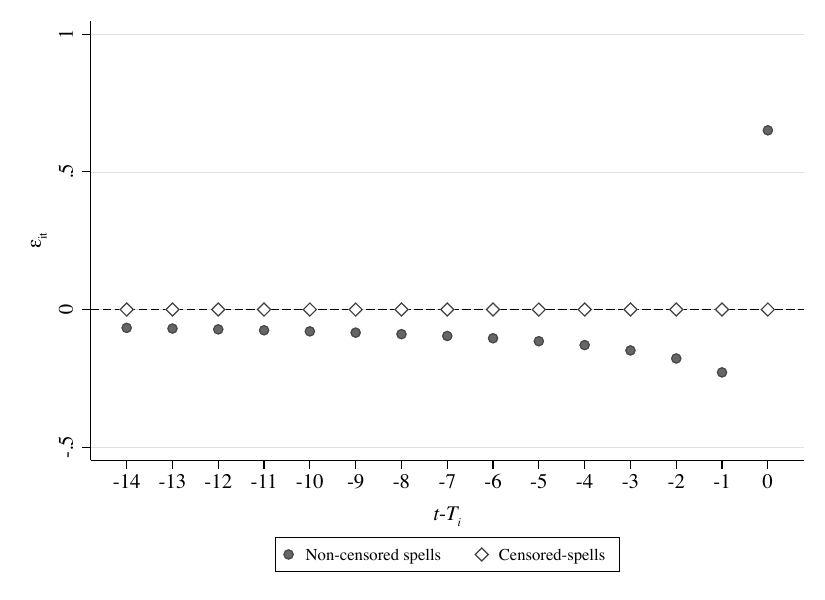}
\end{subfigure}
\begin{subfigure}{\textwidth}
	\caption{$\gamma = -2$}
	\label{fig:1_1_xi_-2}
	\includegraphics[width=0.495\textwidth]{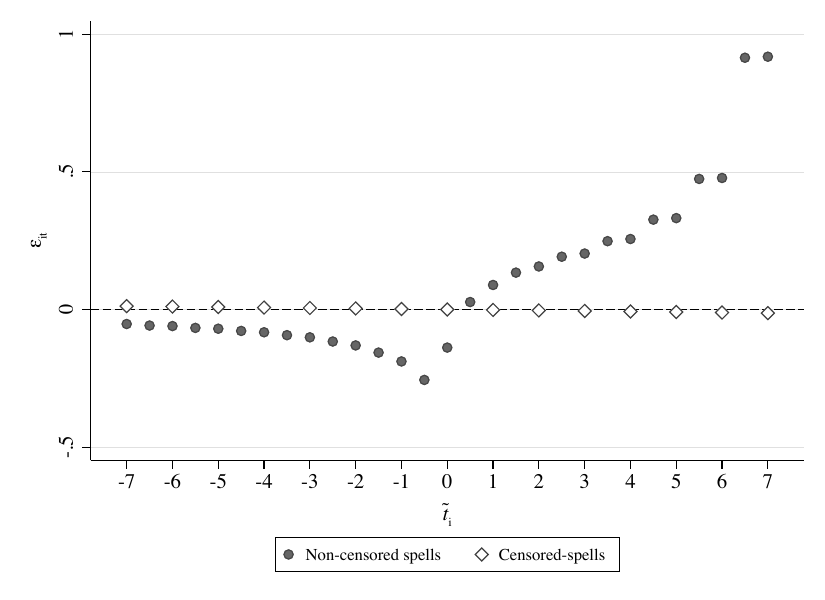}
	\includegraphics[width=0.495\textwidth]{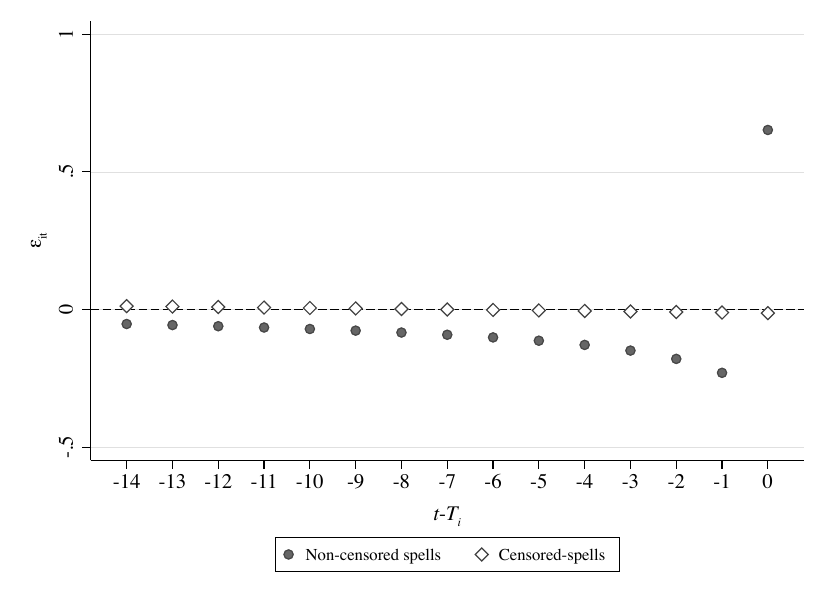}
\end{subfigure}
\ftnotefigure{
	Note: 
	This figure presents evidence on the correlation between the within-individual error term $\tilde\varepsilon_{it}$ and the time dimension, for three different values  of the structural duration dependence parameter $\gamma = 0.05, 0.00$ and $-0.02$.
	For each parameter $\gamma$ value, it reports the average $\tilde\varepsilon_{it} = y_{it} - \lambda_{it}$ per within-time dimension $\tilde t_{i}$ and per duration until the last observation $t-T_i$.
	The graph distinguishes between non-censored and right-censored spells.
	Statistics are obtained from Monte Carlo simulations, with $K =  \DTLfetch{DynamicWriting1}{object}{K}{value}$ repetitions and sample size $N = \DTLfetch{DynamicWriting1}{object}{N}{value}$.
}
\end{figure}

Corresponding results are reported in \autoref{fig:1_2_MC_OLS_FE}. 
As with the average empirical duration profiles, the bivariate model without controls for individual heterogeneity produces biased estimates of the structural duration dependence. 
Including $\alpha_i$ as a control yields estimates that closely track the structural patterns. 
This is not the case for the within-estimation approach, which is commonly advocated to address time-invariant heterogeneity. 
The fixed effects model delivers substantially biased estimates: in all panels, the estimated duration dependence profiles are positive and deviate markedly from their structural counterparts. 
As shown in \autoref{fig:1_1_xi}, these positive \textit{within-estimation duration biases} arise from the positive expected value of $\tilde\varepsilon_{it}$ for positive values of $\tilde t_i$ (or, equivalently, for the last observation $t = T_i = \tau_i$) in the case of non-censored spells.

This simple simulation confirms the existence of a \textit{within-estimation duration bias} when the dependent variable is directly linked to attrition. 
In such cases, the crucial assumption of the linear model is violated, as $\mathbb{E}(\tilde\varepsilon_{it}\,|\,\tilde t_i) > 0$ for non-censored spells. 
Consequently, the structural duration profile recovered by the fixed effects estimator is severely biased, regardless of the true direction of structural duration dependence.



\section{Simulation II: a job search process}


The specification and calibration of the model are specified in \autoref{table:sim_2_calibration}. 
According to those, the conditional expectations of $A_{it}$ and $c_{ait}$ in the first period are characterized as follows : 
\vspace{-10pt}
\begin{align*}
	\mathbb{E}(A_{it}|t = 1) &= \mathbb{E}(\alpha_i) + \beta(1-1) + \mathbb{E}(\xi_{it}) = \frac{\underline \alpha + \overline \alpha}{2} + \frac{\underline \xi + \overline \xi}{2} = 10.5 \\[5pt]
	\mathbb{E}(c_{ait}|t = 1) &= \mathbb{E}_i\left(\mathbb{E}(c_{ait}|t = 1, i) | t = 1\right) = \mathbb{E}_i\left(\mathbb{P}(c_{ait}|t = 1, i) | t = 1\right) \\
	&=
	\mathbb{E}_i\left(\gamma_0 + \gamma_1 \alpha_i + \gamma_2(t-1)  | t = 1\right) =
	\gamma_0 + \gamma_1 \mathbb{E}(\alpha_i)+ \gamma_2(1-1) \\ &= 
	\gamma_0 + \gamma_1 \frac{\underline \alpha + \overline \alpha}{2} = 0.05
\end{align*}

\renewcommand{\arraystretch}{1.5}
\begin{table}[h!]
	\centering
	\caption{
		Simulation II,
		specification and calibration
	}
	\label{table:sim_2_calibration}
	\footnotesize
	\begin{tabularx}{\textwidth}{| X | L{5.0cm} | L{5.0cm} |} 
		\cline{2-3}
		\multicolumn{1}{c|}{} 
		& \textbf{Specification}
		& \textbf{Calibration} \\
		\hline
		\textbf{DGP}
		& $ \phi(\alpha_i, t ; \beta) =  \lfloor \alpha_i + \beta (t-1) + \xi_{it} \rceil $
		& $\beta = -0.20$ \\
		& $\rho_c(\alpha_i, t;\gamma) = \gamma_0 + \gamma_1 \alpha_i + \gamma_2(t-1) $ & $\gamma_0 = \frac{7}{50}$, $\gamma_1 = -\frac{3}{350}$, $\gamma_2 = -\frac{1}{1150}$ \\
		& & $\psi = 0.3$
		\\ \hline 
		\textbf{Heterogeneity}
		& $\alpha_i  \sim \mathcal{U}(\underline{\alpha},\overline{\alpha})$  &
		$\underline{\alpha} = 7$, $\overline{\alpha} = 14$
		\\ \hline 
		\textbf{Error term}
		& $\xi_{it} \sim \mathcal{U}(\underline\xi,\overline\xi)$ 
		& $-\underline \xi = \overline \xi = 1$ 
		\\ 
		\hline 
	\end{tabularx}
	\vspace{10pt}
	\ftnotetable{
		Note: 
		This table reports the specification and calibration used in Simulation II.
	}
\end{table}
\renewcommand{\arraystretch}{1.0}

\begin{figure}[h!]
	\centering
	\caption{Simulation II, dynamic selection}
	\label{fig:1_1_DDJS_MC_alpha_dynamic_selection}
	\includegraphics[width=0.7\textwidth]{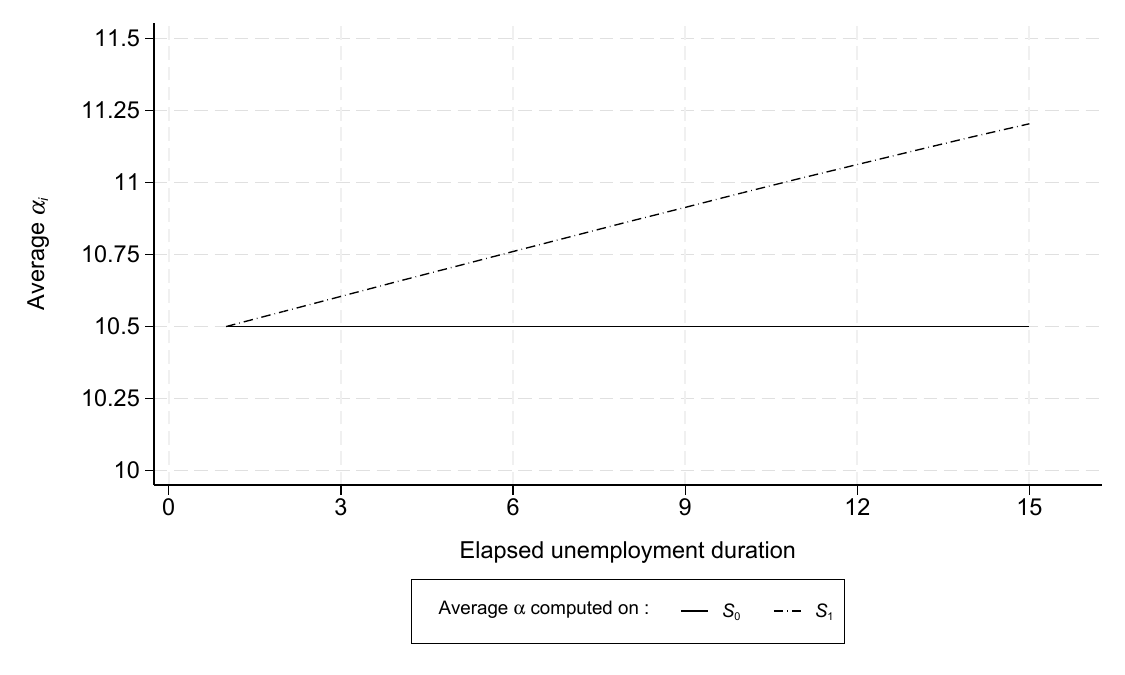}
	\ftnotefigure{
		Note: This figure presents evidence of dynamic selection for Simulation II.
		It reports the average $\alpha_i$ per duration $t$ in the data-generating sample $\mathcal{S}_{0}$ and in the observation sample $\mathcal{S}_{1}$.
		Statistics are obtained from Monte Carlo simulations, with $K =  \DTLfetch{DynamicWriting2}{object}{K}{value}$ repetitions and sample size $N = \DTLfetch{DynamicWriting2}{object}{N}{value}$.
	}
\end{figure}

\begin{figure}[h!]
	\centering
	\caption{Simulation II, linear duration specification}
	\label{fig:linear_u_dur}
	\begin{subfigure}{\textwidth}
		\caption{Applications $A_{it}$}
		\label{fig:linear_A_it}
		\includegraphics[width=1\textwidth, clip, trim = 0cm 0cm 0cm 4.4cm]{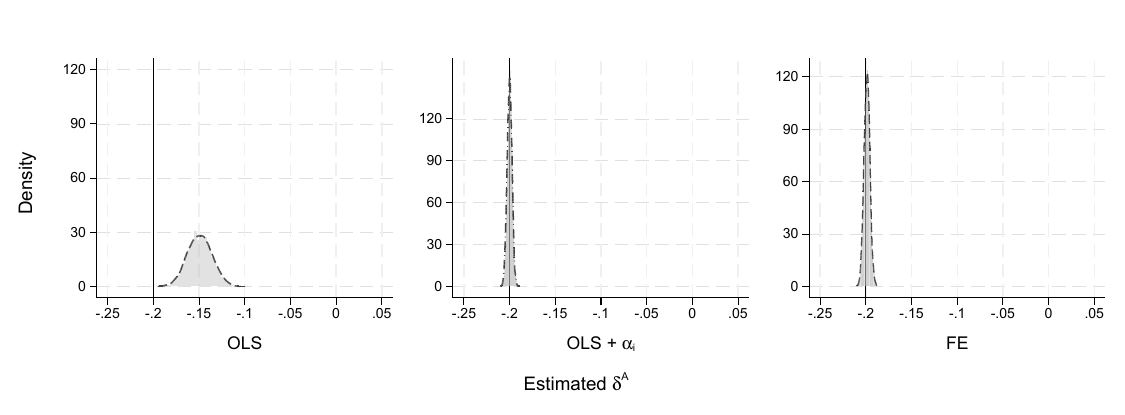}
	\end{subfigure}
	\vspace{10pt}
	
	\begin{subfigure}{\textwidth}
		\caption{Callback rate $c_{ait}$}
		\label{fig:linear_c_ait}
		\includegraphics[width=1\textwidth, clip, trim = 0cm 0cm 0cm 4.4cm]{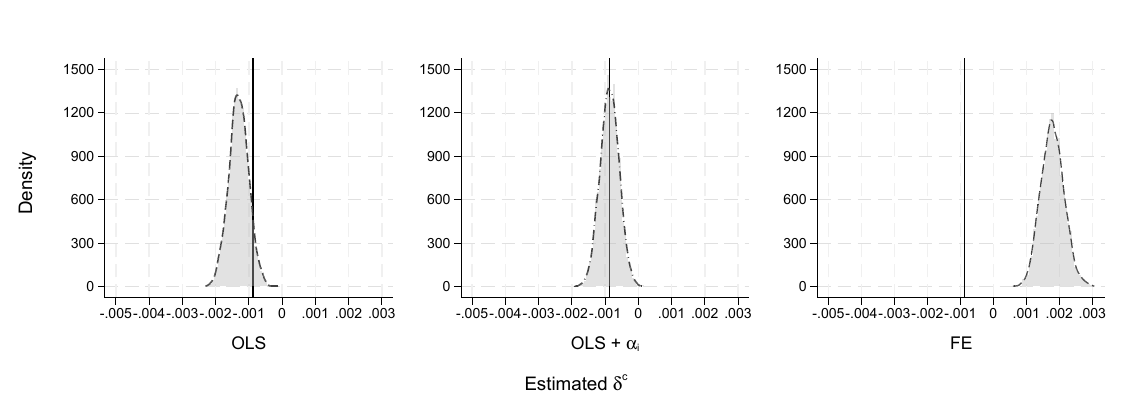}
	\end{subfigure}
	\vspace{10pt}
	
	\begin{subfigure}{\textwidth}
		\caption{Number of callbacks $C_{it}$}
		\label{fig:linear_C_it}
		\includegraphics[width=1\textwidth, clip, trim = 0cm 0cm 0cm 4.4cm]{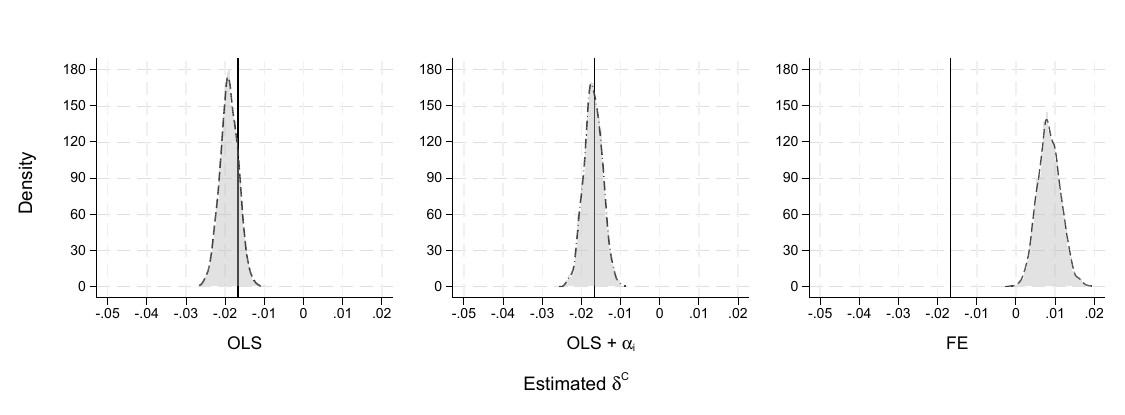}
	\end{subfigure}
	\ftnotefigure{
		Note: 
		This figure reports the distribution of different estimates of structural duration dependence in $A_{it}$, $c_{ait}$ and $C_{it}$.
		The parametric duration functions are specified linearly, \textit{i.e.} $f(t;\delta) = \delta t$.
		The ``structural'' values of $\delta$ are obtained by regressing the dependent variables on time $t$, based on the data generating samples $\mathcal{S}_0$  (without attrition).
		The average of those ``structural'' coefficients across all replications $K$ are reported as solid lines.
		Estimated duration profiles are all based on the observation sample $\mathcal{S}_1$ and are obtained by regressing the outcome on (1) time $t$ only (OLS), (2) time $t$ and $\alpha_i$ (OLS + $\alpha_i$) or (3) time $t$ and a set of individual fixed effects (FE).
		Statistics are obtained from Monte Carlo simulations, with $K =  \DTLfetch{DynamicWriting2}{object}{K}{value}$ repetitions and sample size $N = \DTLfetch{DynamicWriting2}{object}{N}{value}$.
	}
\end{figure}

\begin{figure}[h!]
	\centering
	\caption{Simulation II, control for proxies of $\alpha_i$}
	\label{fig:simulation_2_noisy_alpha}
	\begin{subfigure}{0.495\textwidth}
		\caption{Applications $A_{it}$}
		\label{fig:1_1_DDJS_MC_saturated_A_it_noisy_alpha}
		\includegraphics[width=\textwidth]{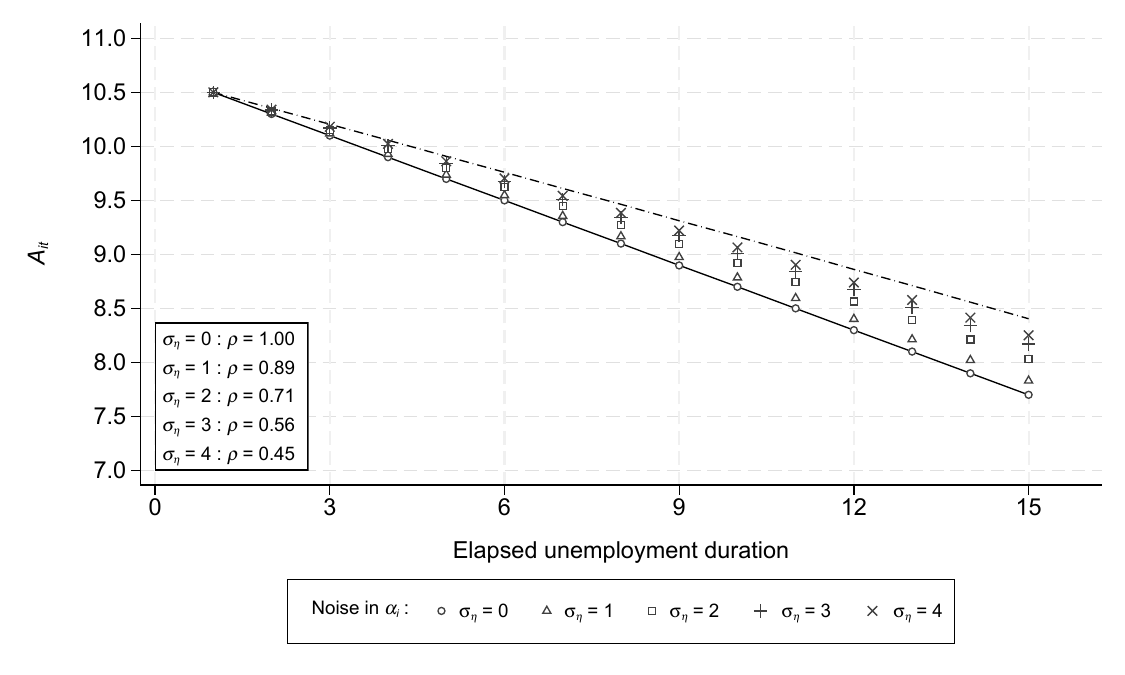}
	\end{subfigure}
	\begin{subfigure}{0.495\textwidth}
		\caption{Callback rate $c_{ait}$}
		\label{fig:1_1_DDJS_MC_saturated_c_ait_noisy_alpha}
		\includegraphics[width=\textwidth]{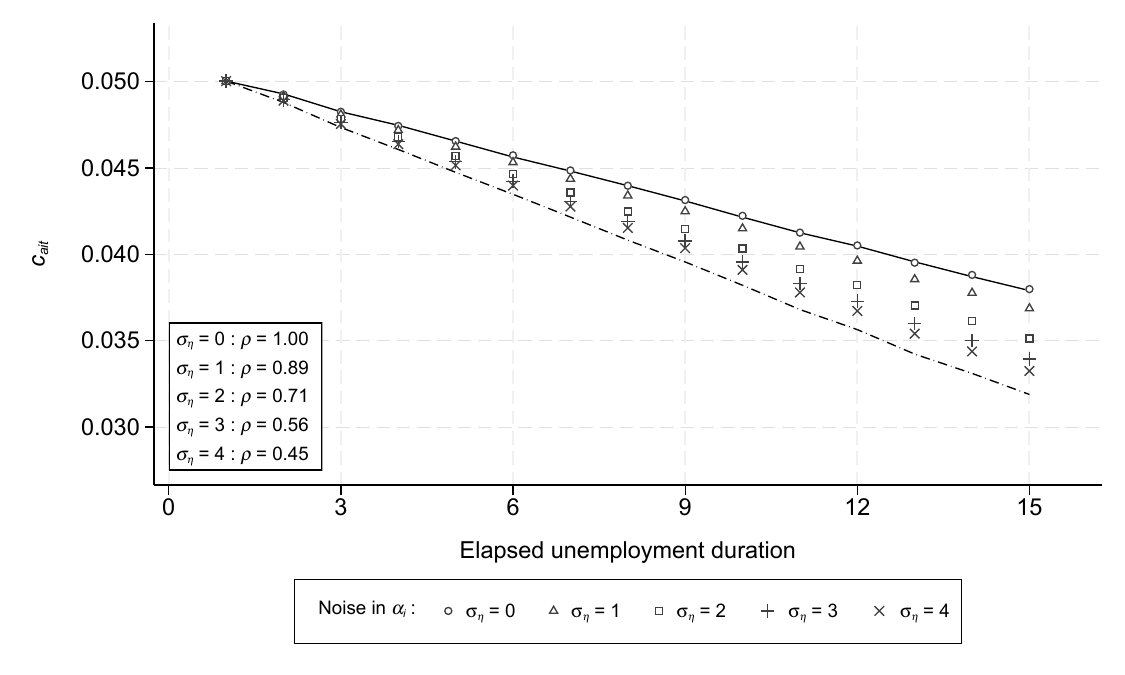}
	\end{subfigure}
	\begin{subfigure}{0.495\textwidth}
		\caption{Number of callbacks $C_{it}$}
		\label{fig:1_1_DDJS_MC_saturated_C_it_noisy_alpha}
		\includegraphics[width=\textwidth]{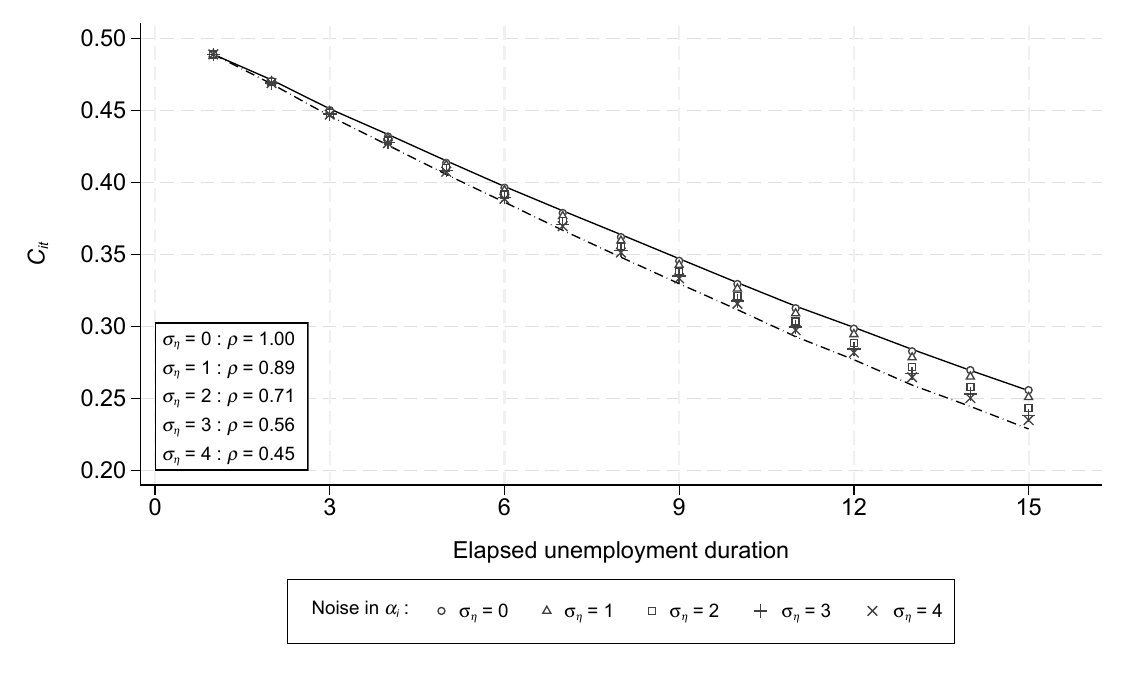}
	\end{subfigure}
	\ftnotefigure{
		Note: 
		This figure reports empirical estimates of structural duration dependence in $A_{it}$, $c_{ait}$ and $C_{it}$.
		The structural duration profiles to be estimated are depicted as solid lines, while the empirically observed duration profiles are reported as dashed lines.
		The parametric duration functions $f(t;\delta)$ are specified in a saturated manner, \textit{i.e.} $f(t;\delta) = \{\delta_t\}_{1..\bar\tau}$.
		Estimated profiles are all based on the observation sample $\mathcal{S}_1$ and are obtained by regressing the outcomes on time $t$ and a (potentially) noisy proxy of heterogeneity, $\tilde \alpha_i = \alpha_i + \mathcal{N}(0,\sigma_\eta)$, whose correlation with the true $\alpha_i$ is reported on the graph.
		Statistics are obtained from Monte Carlo simulations, with $K =  \DTLfetch{DynamicWriting2}{object}{K}{value}$ repetitions and sample size $N = \DTLfetch{DynamicWriting2}{object}{N}{value}$.
	}
\end{figure}

\begin{figure}[h!]
	\centering
	\caption{Simulation II, correlation between $\varepsilon$, $\tilde t_{i}$ and $t-\tau_i$}
	\label{fig:1_2_xi}
	\begin{subfigure}{\textwidth}
		\caption{Applications $A_{it}$}
		\label{fig:1_1_xi_A_it}
		\includegraphics[width=0.495\textwidth]{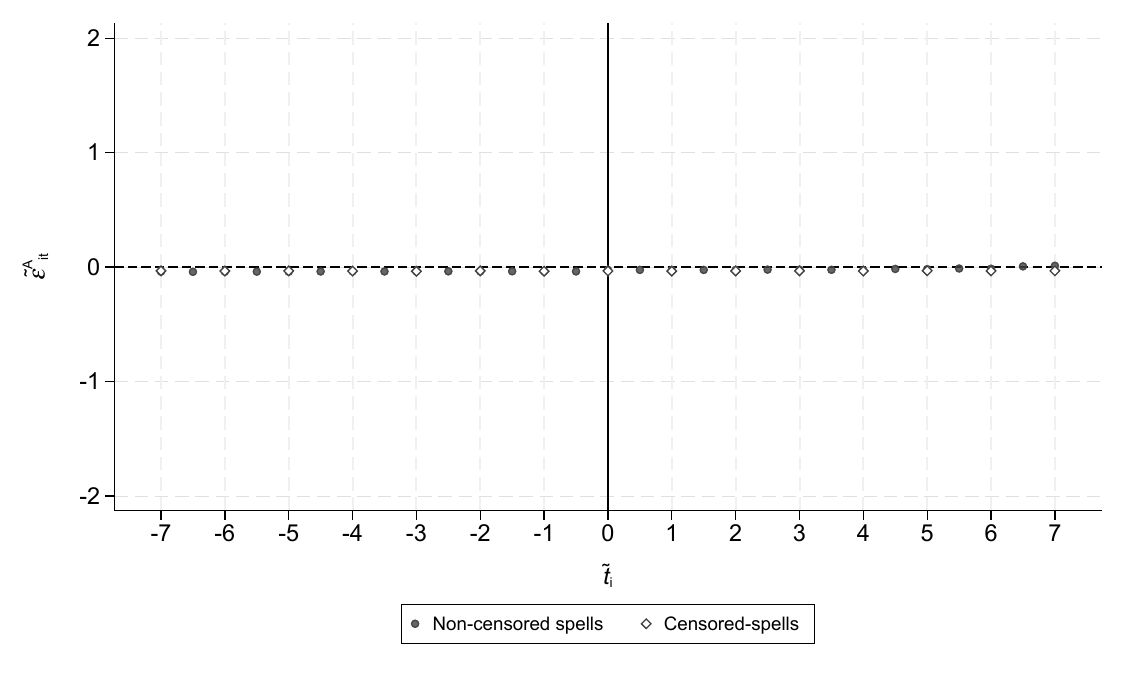}
		\includegraphics[width=0.495\textwidth]{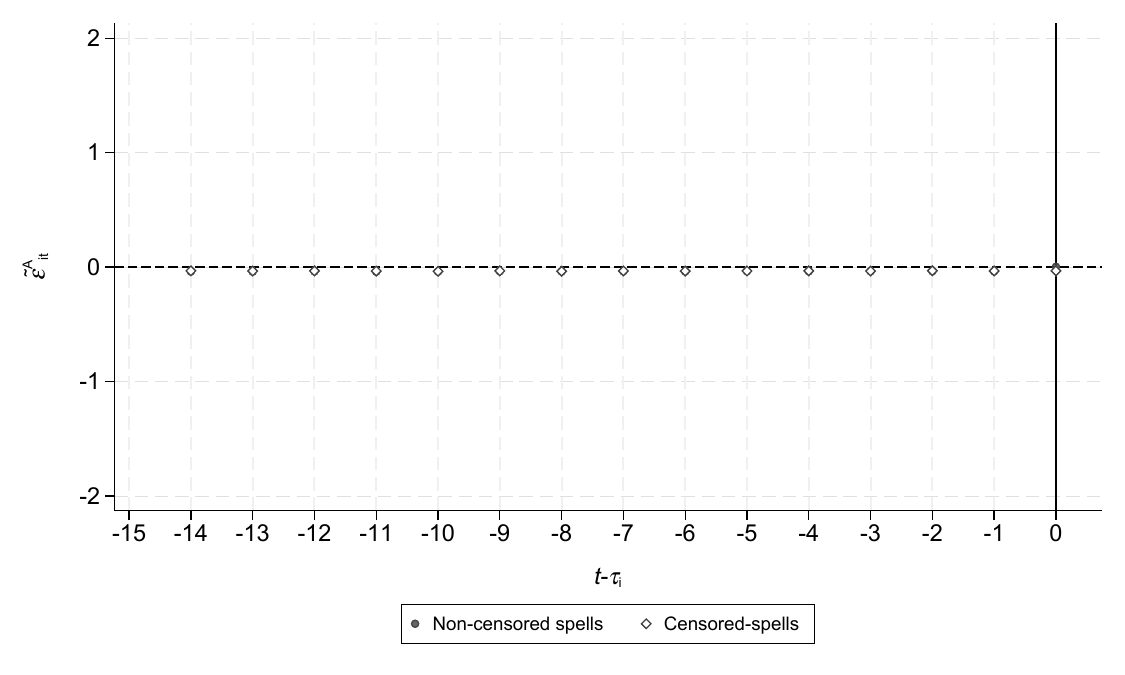}
	\end{subfigure}
	\begin{subfigure}{\textwidth}
		\caption{Callback rate $c_{ait}$}
		\label{fig:1_1_xi_c_ait}
		\includegraphics[width=0.495\textwidth]{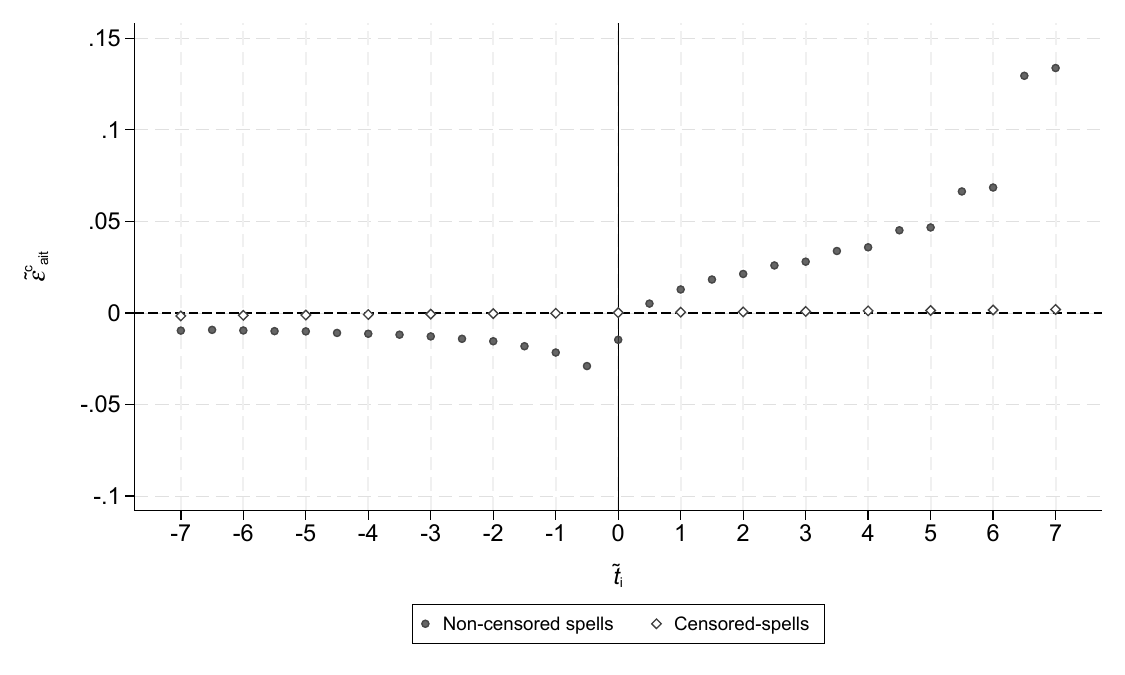}
		\includegraphics[width=0.495\textwidth]{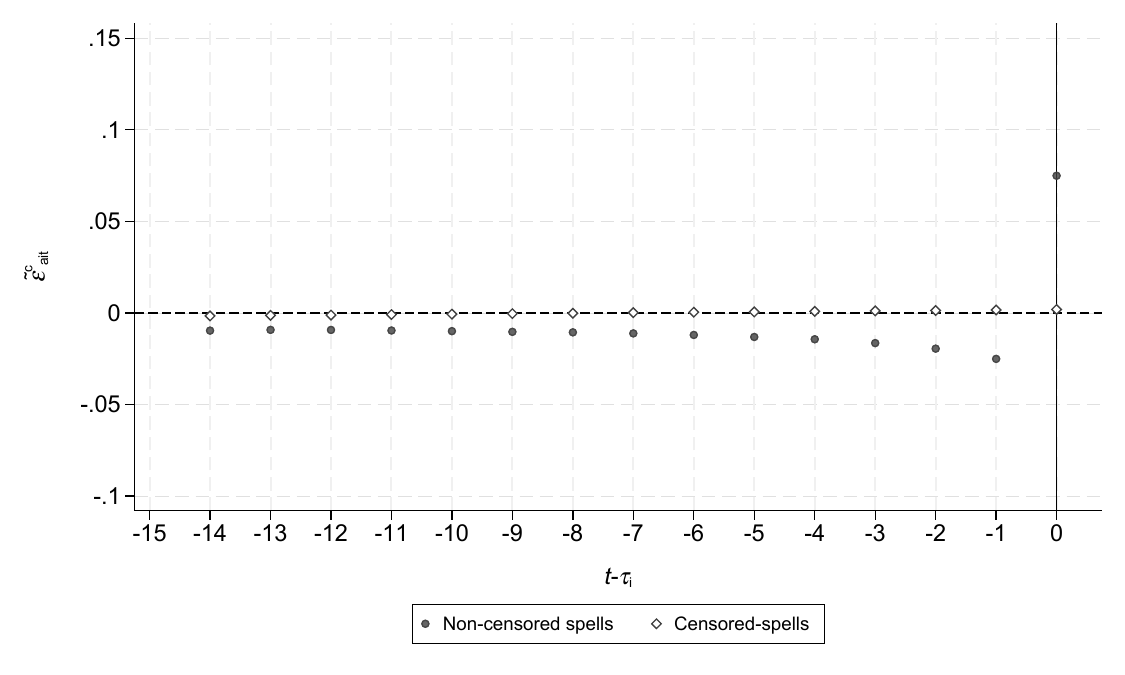}
	\end{subfigure}
	\begin{subfigure}{\textwidth}
		\caption{Callbacks $C_{it}$}
		\label{fig:1_1_xi_C_it}
		\includegraphics[width=0.495\textwidth]{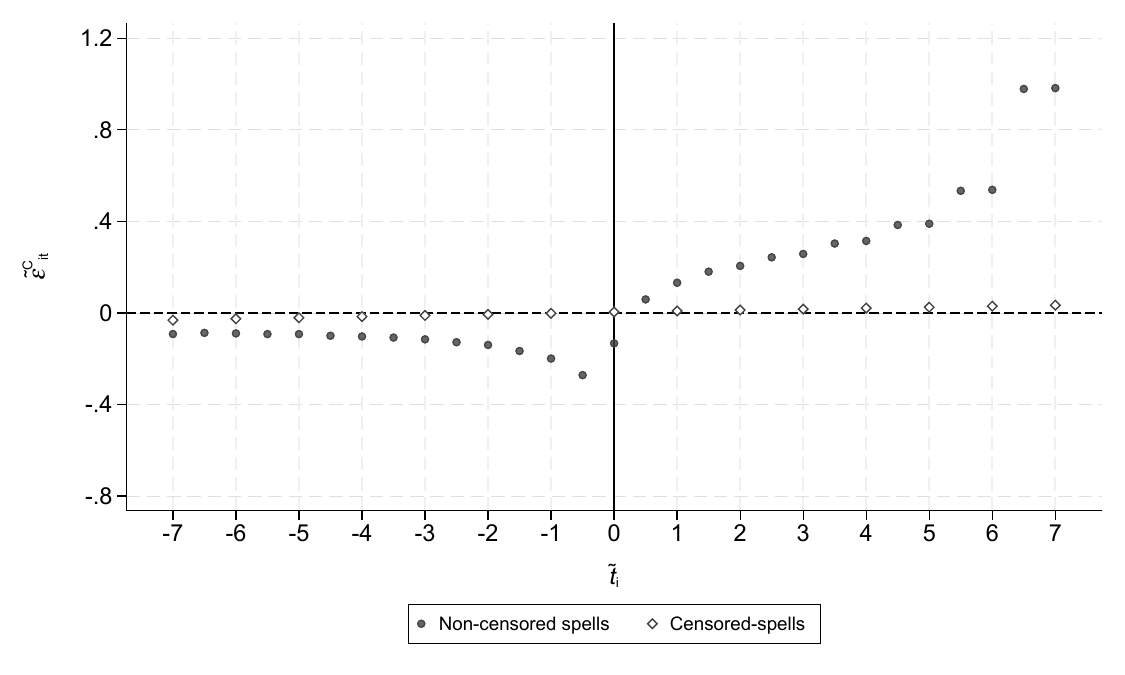}
		\includegraphics[width=0.495\textwidth]{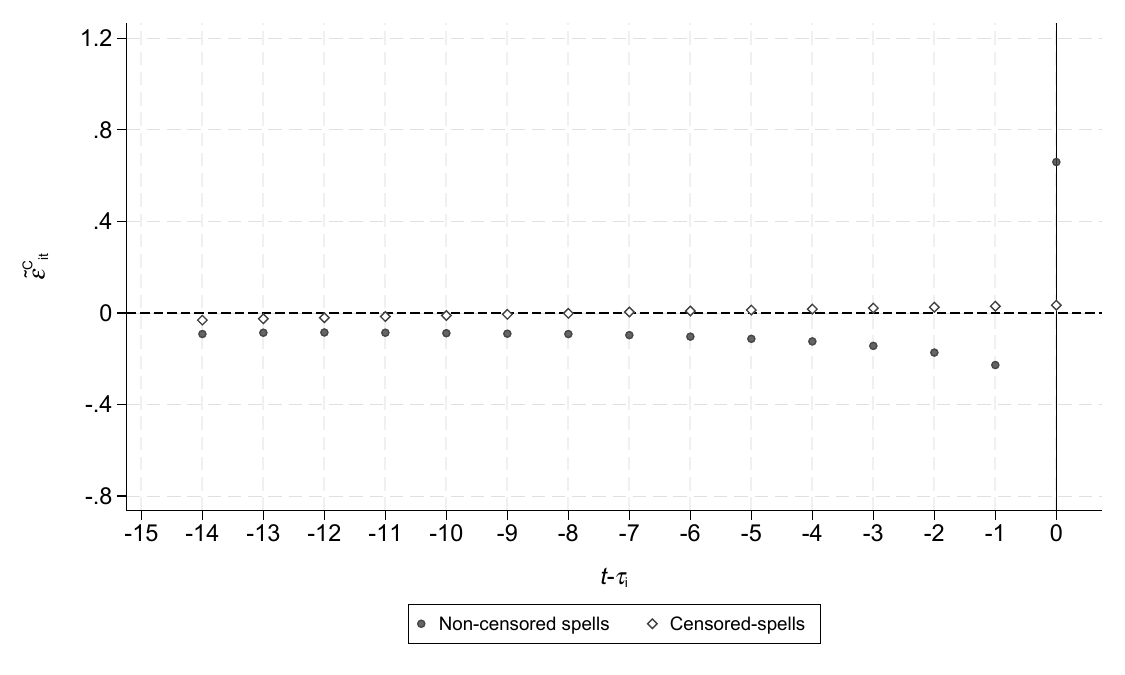}
	\end{subfigure}
	\ftnotefigure{
		Note: 
		This figure presents evidence on the correlation between the within-individual error term $\tilde\varepsilon_{it}$ and the time dimension, for variables $A_{it}$, $c_{ait}$ and $C_{it}$.
		For each variable, it reports the average $\tilde\varepsilon_{it}$ per within-time dimension $\tilde t_{i}$ and per duration until the last observation $t-T_i$.
		The graph distinguishes between non-censored and right-censored spells.
		Statistics are obtained from Monte Carlo simulations, with $K =  \DTLfetch{DynamicWriting2}{object}{K}{value}$ repetitions and sample size $N = \DTLfetch{DynamicWriting2}{object}{N}{value}$.
	}
\end{figure}

\end{document}